%% file: DeRiskDB.tex
\documentclass[preprint,12pt]{elsarticle} 

\usepackage{graphicx}
\usepackage{amssymb}
\usepackage{amsmath}
\usepackage{lineno}

\usepackage{rotating}

\usepackage{subcaption}
\RequirePackage{doi}
\usepackage{hyperref}
\usepackage{todonotes}
\presetkeys{todonotes}{inline}{}
\usepackage{natbib}
\usepackage{breqn}
\usepackage{booktabs}
\usepackage[utf8]{inputenc}
\usepackage{multicol}
\usepackage{xfrac}
\usepackage{mathtools}
\usepackage{fontawesome5}

\providecommand{\e}[1]{\ensuremath{\cdot 10^{#1}}}

\journal{Marine Structures}

\begin{document}

\begin{frontmatter}


\title{The DeRisk database: Extreme Design Waves for Offshore Wind Turbines}



\author[wind]{Fabio Pierella \corref{cor}}
\author[mek]{Ole Lindberg}
\author[wind]{Henrik Bredmose}
\author[mek]{Harry Bingham}
\author[mek]{Robert W. Read}
\author[compute]{Allan P. Engsig-Karup}

\address[wind]{Department of Wind Energy, DTU }
\address[mek]{Department of Mechanical Engineering, DTU}
\address[compute]{Department of Applied Mathematics and Computer Science, DTU}
\cortext[cor]{Corresponding author \href{mailto:fabpi@dtu.dk}{fabpi@dtu.dk} \\
This work is under the CC-BY-NC-ND 4.0 license \faCreativeCommons\ \faCreativeCommonsBy\ \faCreativeCommonsSa}

\begin{abstract}
The estimation of extreme loads from waves is an essential part of the design of an offshore wind turbine.
Standard design codes suggest to either use simplified methodologies based on regular waves, or to perform fully nonlinear computations. The former might not provide an accurate representation of the real extreme waves, while the latter is computationally too intensive for fast design iterations.
Here, we address these limitations by using the fully nonlinear solver OceanWave3D to establish the DeRisk database, a large collected dataset of extreme waves kinematics in a two-dimensional domain. From the database, which is open and freely available, a designer can easily extract fully-nonlinear wave kinematics for a wave condition and water depth of interest by identifying a suitable computation in the database and, if needed, by Froude-scaling the kinematics.

The fully nonlinear solver is first validated against the DeRisk model experiments at two different water depths, $33.0 [m]$ and $20.0 [m]$, and an excellent agreement is found for the analyzed cases. The experiments are used to calibrate OceanWave3D's numerical breaking filter constant, and the best agreement is found for $\beta=0.5$.
We then compare the experimental static force with predictions obtained by the DeRisk kinematics database and the Rainey force model, and with state-of-the-art industrial practices. For milder storms, we find a good agreement in the predicted extreme force between the present methodology and all of the standard methodologies. At the deep location and for stronger storms, the largest loads are given by slamming loads due to breaking waves. In this condition, the database methodology is less accurate than the embedded stream function method and more accurate than the WiFi JIP methodology, although providing generally nonconservative estimates.
For strong storms at the shallower location, where wave breaking is less dominating, the database methodology is the most accurate overall.

\end{abstract}

\begin{keyword}
wave \sep database \sep wave kinematics \sep nonlinear waves \sep monopile \sep force model \sep breaking wave \sep potential flow \sep offshore wind turbines \sep foundation \sep DeRisk


\end{keyword}

\end{frontmatter}

% \tableofcontents
% \newpage


%% INTRODUCTION
\input{texFiles/intro.tex}

%% METHODOLOGY
\input{texFiles/meth.tex}

%% PARAMETER SPACE
%% \input{texFiles/parspace.tex}

%% RESULTS
\input{texFiles/res.tex}

%% CONCLUSIONS
\input{texFiles/conc.tex}

\section{Online availability}
The DeRisk database is freely available at \url{https://data.dtu.dk/articles/The_DeRisk_Database/10322033}. In the event of publication of work resulting from the use of the model, appropriate referencing to the dataset \cite{Pierella2020database} and this paper should be included.

\section{Acknowledgements}

This work was performed in the framework of the Danish Innovation Fund project
DeRisk, grant no. 4106-00038B.


%% References


\bibliographystyle{model1-num-names}
\bibliography{Fabio}






% \end{thebibliography}

\end{document}

%% file: texFiles/intro.tex
\section{Introduction}
\label{sec:intro}

The substructures for offshore wind turbines are a central element for targeted cost reductions in offshore wind energy. As rotor dimensions increase, the substructures become larger, and extreme loads from storm waves can become design drivers. Therefore, the industry needs accurate and reliable (de-risked) design methodologies for extreme waves. 
At present, these modelling methods fall into two camps. Either they are fast but have limited accuracy, or they are accurate but too computationally demanding for everyday design use.
The main idea of this paper is to bridge this divide by establishing an extensive and openly accessible database of two-dimensional (2D) nonlinear wave kinematics, that can be used to estimate Ultimate Load States (ULS) on offshore wind turbines via integrated load modelling tools.

The IEC standards \cite{iec61400-1} presently define two different methodologies to compute nonlinear waves for extreme load evaluation.
In the constrained wave approach a regular stream function wave is embedded into a background irregular linear wave realization, to account for the nonlinearity of the largest wave.
For example, \citet{Rainey2007} proposed to condition the background linear sea state via the New Wave theory by \citet{Tromans1991} to generate a large, linear wave at a predetermined time and location, that can be suitably replaced by a nonlinear wave. \citet{Pierella2017} use the Hilbert transform to identify the embedment location and the parameters of the embedded stream function wave.

While these methodologies are intuitive and fast, they have some shortcomings. 
First, the largest static load is not always associated with the largest free surface elevation peak. Smaller waves can break, generating higher loads than larger, nonbreaking waves. When the structure is inertia dominated, the force is in the first approximation proportional to the wave steepness, and the large wave is not necessarily the steepest.
Secondly, real-life extreme waves usually propagate with no constant form, have nonsymmetric crests, and tend to break, violating the assumptions of stream-function theory, which only apply to symmetric regular waves on a flat bed.
Moreover, to specify the height of the embedded wave, a Rayleigh distribution is assumed, and the significant wave height $H_S$ is multiplied by a factor, e.g. 1.86 for a 3-hour (hr) realization. However the crest statistics for large nonlinear sea states do not necessarily follow a Rayleigh distribution. Lastly, especially for flexible structures, the maximum load is not necessarily associated with the largest waves \cite{Bredmose2013b} as it is a random combination of static and dynamic loads. 

In the second suggested approach, fully nonlinear computations for a given sea state are used to assess the loads. Here, all waves are nonlinear and assumptions regarding the relationship between the free-surface elevation and the maximum induced loads are not required.
Following this approach, \citet{Paulsen2013} compute the loads on a stiff cylinder due to an irregular sea state on a slope via two different nonlinear methods. In the first, wave kinematics are produced via the fully nonlinear potential flow solver OceanWave3D (OW3D) \cite{Engsig-Karup2009}, and the loads on the structure are computed via a slender body force model. In the second, the Navier-Stokes solver OpenFoam \cite{Jasak2007} is run in a domain where the cylinder surface is directly modelled. While the two methods were able to accurately reproduce the experimental horizontal force, they respectively require 20 minutes on a single-core machine and several weeks on a computational cluster to generate a 100-second time series. 
It is impractical to use fully-nonlinear methods of this type to examine the many load cases considered during the design of the substructure.

In the current method, developed in the framework of the DeRisk Innovation project \cite{Bredmose2016}, we can retain the full nonlinearity and minimize the time needed to obtain a certain wave realization by precomputing an extensive database of extreme wave kinematics, eliminating the need for an individual to perform fully nonlinear simulations. The database is freely available in an online repository \cite{Pierella2020database}. Dedicated software is provided to access the time series corresponding to the relevant site conditions, and to perform Froude-scaling where necessary.
% Designers can extract time series of wave kinematic from the database without having to perform the simulations themselves.
With this paper we describe the calculation and compilation of the database, we validate and calibrate the model constants against lab-scale experimental data, and finally we compare the current methodology with state-of-the-art industrial practices.

%% file: texFiles/meth.tex
\section{Methodology}
\label{sec:meth}

We generate nonlinear kinematics for a large number of extreme sea states, with combinations of significant wave heights $H_S$, peak periods $T_P$ and water depths $h$, that are representative for storms in offshore locations.

With these three parameters and the constant gravity $g$, it is possible to define two nondimensional numbers that completely define the parameter space

\begin{multicols}{2}
\begin{equation}
    h^*=\frac{h}{gT_P^2} \label{eq:ndh}
\end{equation}\break
\begin{equation}
    H_S^*=\frac{H_S}{gT_P^2} \label{eq:ndhs}  
\end{equation}
\end{multicols}

Following standard practices \cite{veritas2007recommended}, the peak enhancement factor $\gamma$ was

\begin{align}
  \gamma = 
  \begin{cases}
    5 & T_P/\sqrt{H_S} \le 3.6 \\
    \text{exp}(5.75-1.15T_P/\sqrt{H_S}) & 3.6 < T_P/\sqrt{H_S} \le 5 \\
    1 & 5 < T_P/\sqrt{H_S} \\
  \end{cases}  
\end{align}

\begin{figure}
    \centering
    \includegraphics[width=\textwidth]{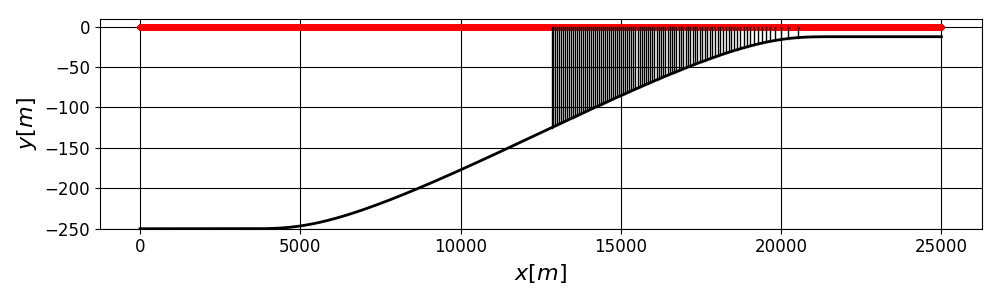}
    \caption{Computational domain. The red line indicates the still water line. The vertical black lines indicate the locations at which the kinematics was sampled. The thick black line is the sea bottom.}
    \label{fig:compDom}
\end{figure}

To generate the nonlinear kinematics, we create a two-dimensional (2D) computational domain on which we execute the nonlinear potential flow model OceanWave3D \cite{Engsig-Karup2009}. The domain is specified in \autoref{fig:compDom}.

In order to generate independent runs, we choose 8 different values of the nondimensional number $H_S^*$, see \autoref{tab:database_sea_states}.
For each sea state, different realizations are run. For each realization, a different set of random phases in the interval $[0,2\pi)$ is sampled from a uniform distribution and assigned to each wave component of the generated JONSWAP spectrum. 
A total of 67 different surface elevation signals were imposed in a relaxation zone at the offshore boundary of the OW3D domain. As each simulation ran, waves propagated from the deep offshore region to the shallow onshore region of the domain.
To collect results at different nondimensional depths $h^*$, the wave kinematics were sampled at the shallow part of the domain along vertical lines between the bed and the free surface at 144 different $x$-locations. In \autoref{fig:compDom}, each sampling location is highlighted with a black line. Further details on the computational domain and on the solver are given in the following sections.

\begin{table}
  \centering
  \begin{tabular}{c|ccc|cc}
    \toprule
    & $H_S$ & $T_P$   & $\gamma$  & $H_S/{gT_P^2}$ & $N_R$\\
    \midrule
    1   & 4.5   &  15.15  &  1     &  2.0\e{-3} & 9 \\    
    2   & 6.76  &  15.15  &  1     &  3.0\e{-3} & 9\\ 
    3   & 9.01  &  15.15  &  1     &  4.0\e{-3} & 7 \\ 
    4   & 11.26 & 15.15   &  1.75  &  5.0\e{-3} & 8 \\
    5   & 13.51 & 15.15   &  2.75  &  6.0\e{-3} & 9 \\
    6   & 15.77 & 15.15   &  3.9   &  7.0\e{-3} & 8 \\
    7   & 18.02 & 15.15   &  5     &  8.0\e{-3} & 10 \\
    8   & 22.52 & 15.15   &  5     &  1.0\e{-2} & 7 \\
    \midrule
    Total   & & & & & 67 \\
  \bottomrule
  \end{tabular}
  \caption{Sea states input in the computational domain. $N_R$ is the number of realizations that were run for each sea state.}
  \label{tab:database_sea_states}
\end{table}

\subsection{The nonlinear potential solver OceanWave3D}

The OceanWave3D flow model was developed at DTU as a robust and fast fully-nonlinear potential solver \cite{Engsig-Karup2009}.
The code solves the Laplace equation by a high-order finite difference method, taking into account the fully nonlinear free surface boundary kinematic and dynamic conditions.

\begin{align}
  \partial_t\eta = -\nabla \eta \cdot \nabla \tilde{\phi} + \tilde{w}(1+\nabla \eta \cdot \nabla \eta) \\
  \partial_t \tilde{\phi} = -g\eta - \frac{1}{2}(\nabla\tilde{\phi} \cdot \nabla\tilde{\phi} - \tilde{w}^2(1+\nabla \eta \cdot \nabla \eta),
  \label{eq:laplaceproblemxyz}
\end{align}

where $\partial_x \psi  = \partial \psi/ \partial x$, and the tilde stands for quantities evaluated at the free surface. In the above equation, $\eta$ is the free surface elevation, $\phi$ is the fluid scalar potential, $t$ is time, and $(u,v,w)$ are the components of the velocity vector in the $(x,y,z)$ directions.

The velocity potential satisfies

\begin{align}
  \phi = \tilde{\phi}, & \quad z=\eta \\
  \nabla^2 \phi + \partial_{zz}\phi = 0, & \quad -h\le z<\eta \\
  \partial_z\phi + \nabla h \cdot \nabla \phi =0, & \quad z=-h \label{eq:bottom}
\end{align}

\autoref{eq:bottom} is the condition of impermeability of the seabed.

In the OceanWave3D formulation, the $(x,y,z)$ coordinate system is changed to a $(x,y,\sigma)$ system, which makes it possible to solve the problem in a time-invariant grid

\begin{align}
  \sigma \equiv \frac{z+h(x,y)}{\eta(x,y,t)+h(x,y)} \in [0,1]
\end{align}

Once the potential is solved for, the velocity is obtained via differentiation in different directions. Further details are available in \citet{Engsig-Karup2009}.

\subsubsection{The numerical domain}
\label{sec:numerical_domain}

The domain is discretized by $(N_x,N_y,N_Z) = (16385, 1, 17)$ points.
The domain is $L_x=25\e 3 [m]$ long, which resulted in a grid size of $\Delta x = 1.53[m]$.
The depth $L_z$ varies from $h=250[m]$ to $h=12.5[m]$ at the outlet. 

The average slope is $1/100$, with a tangent hyperbolic shape, as defined by

\begin{flalign}
  x^*=\frac{x-L_{\text{in}}}{L_{\text{out}}-L_{\text{in}}}-\frac{1}{2}, x \in [0, L_x] \\
  h(x)=
  \begin{dcases}
      h_{\text{in}}  & x^*\le -\frac{1}{2}  \\
      h_{\text{in}}-\frac{1}{2}(h_{\text{in}}-h_{\text{out}})\left(1+\tanh\frac{\sin (\pi x^*)}{1-4x^{*2}}\right), & -\frac{1}{2}  <x^*\le \frac{1}{2} \\ 
      h_{\text{out}} & x^* > \frac{1}{2}  \\
  \end{dcases}
  \label{eq:seabottom}
  \end{flalign}

  The wave generation zone is positioned between $x=0$ and $x=L_{\text{in}}$, while the wave absorption zone is between $x=L_{\text{out}}$ and $x=L_x$. For the current domain, the following set of parameters is used

\begin{align*}
  h_{\text{in}}=250.0[m], h_{\text{out}} = 12.5[m] \\
  L_{\text{in}}=2.5 \e 3[m], L_{\text{out}}=22.5 \e 3[m], L_x=25.0 \e 3[m]
  \label{eq:domain_parameters}
\end{align*}

 The spectrum high-cut (hc) frequency is set at $f_{\text{hc}}=1/3 [Hz]$ for all of the performed simulations, corresponding to a wave with linear wavelength of $L_{\text{hc}}=14.04[m]$ and $kh_{\text{hc}}=110.39$.
Thus, the shortest wave in the spectrum is resolved with a minimum of $n_x=9$ grid points in the $x$-direction. For every grid, we indicate with $N_i$ the total number of points in the direction $i$, and with $n_i$ the number of points per wavelength in the direction $i$.
The input spectrum is discretized into $2^{19}$ components in the frequency axis, to achieve a $\Delta t = 0.07 [s]$, which guarantees a return time of 10 hours. Each simulation is run for 8 hours of physical time. The first 1.7 hours are left out to allow the shortest generated wave in the linear spectrum to travel to the last of the sampling locations. 

\subsubsection{Wave generation and wave absorption strategy}

OceanWave3D implements different wave generation techniques. Among others, it is possible to impose linear irregular waves in a relaxation zone or to generate waves by enforcing a boundary flux. 
To generate the database, a $L_{\text{in}}=2500[m]$ long wave generation zone is employed. Linear wave theory is used to calculate the prescribed linear solution for all of the sea states specified in \autoref{tab:database_sea_states}.
At the outlet of the domain, a numerical beach absorbs the outgoing waves. OceanWave3D implements a linear friction damping applied to the tangential velocity, following the strategy by \citet{Clamond2005}, which dissipates energy and minimizes reflections.
In the current computations, the wave absorption zone is $2500[m]$ long.

\subsubsection{Breaking filter}

% A potential flow solver cannot directly simulate breaking, which is a viscous phenomenon.
% However, the reproduced waves can get very close to the breaking limit. When this occurs, if energy is not subtracted from the domain, the free surface can become discontinuous and the solution becomes unstable. 
To prevent instabilities arising from near-breaking waves, OceanWave3D has a simple breaking filter that smooths the free surface when the downward Lagrangian particle acceleration ${Dw}/{Dt}$ overcomes a certain fraction $\beta$ of the gravitational acceleration $g$. If the downward acceleration is lower than the threshold at a domain location, the solver smooths a 10-point region centered at the point with a 3-point filter, where the $i^{th}$ free surface elevation is computed as

\begin{align}
  \eta_i = 0.25\eta_{i-1} + 0.5\eta_i + 0.25\eta_{i+1}.
\end{align}

The parameter $\beta$ is typically $\beta \in [0.3,0.7]$. Physically, if a particle at the free surface is falling with an acceleration of $g$ it is in free fall, and therefore the wave is breaking. From previous studies, we observed that a value of $\beta=0.5$ usually gives good agreement with the experiments \cite{schloer2017experimental, Ghadirian2019c}. An investigation into the choice of the $\beta$ factor is given in \autoref{sec:results}.

\subsubsection{Grid convergence study}

In \citet{Engsig-Karup2009}, a detailed discussion of the accuracy and convergence of the solver is provided. The authors investigated the effect of the number of vertical points $N_z$ on the dispersion error for a linear wave, in combination with the size of the differentiation scheme $r$, the clustering of the grid towards the surface and the number of grid points per wavelength $n_x$.
% By choosing the spatial resolutions one can essentially control the accuracy of the simulations, and in aprticular this allows for tuning the dispersion properties to the problem at hand.
% For one-dimensional first- and second-order derivatives in the $x$, $y$ and $\sigma$ direction, a discretization scheme of width $r$ leads to an accuracy of $(r-1)^{th}$ order in the differentiation. 

When $r=7$ and $N_z=17$ clustered points are used in the vertical direction, \citeauthor{Engsig-Karup2009} show that the phase error on a linear wave is close to zero up to a $kh=30.0$, provided $n_x=9$ points per wavelength are employed in the horizontal direction. They conclude that using a clustered grid in the vertical direction brings great advantages over an evenly spaced grid in terms of convergence, mainly when a higher-order scheme is used for the differentiation.

A similar conclusion was drawn by \citet{Schloer2013}, who performed a thorough analysis of the necessary discretization to describe an irregular sea state on a shoaling domain, similar to the one that was analyzed in the current work.
The shortest wave in their computational domain had a frequency of $f=0.50[Hz]$ and a wavelength of $\lambda=6.25[m]$, which led to a $kh\approx 80$ in the deepest side of the domain. They found that a grid with $N_z=17$ points in the vertical direction and $n_x=8$ was necessary to achieve a good convergence of the simulation, both for the phase and the amplitude dispersion.
Based on the above discussion, $N_x=16835$ grid points and $N_z=17$ were chosen for the production grid of the current simulations.

\begin{figure}[hbtp!]
  \centering
  \includegraphics[width=\textwidth, clip, trim={0 7.5cm 0 0}]{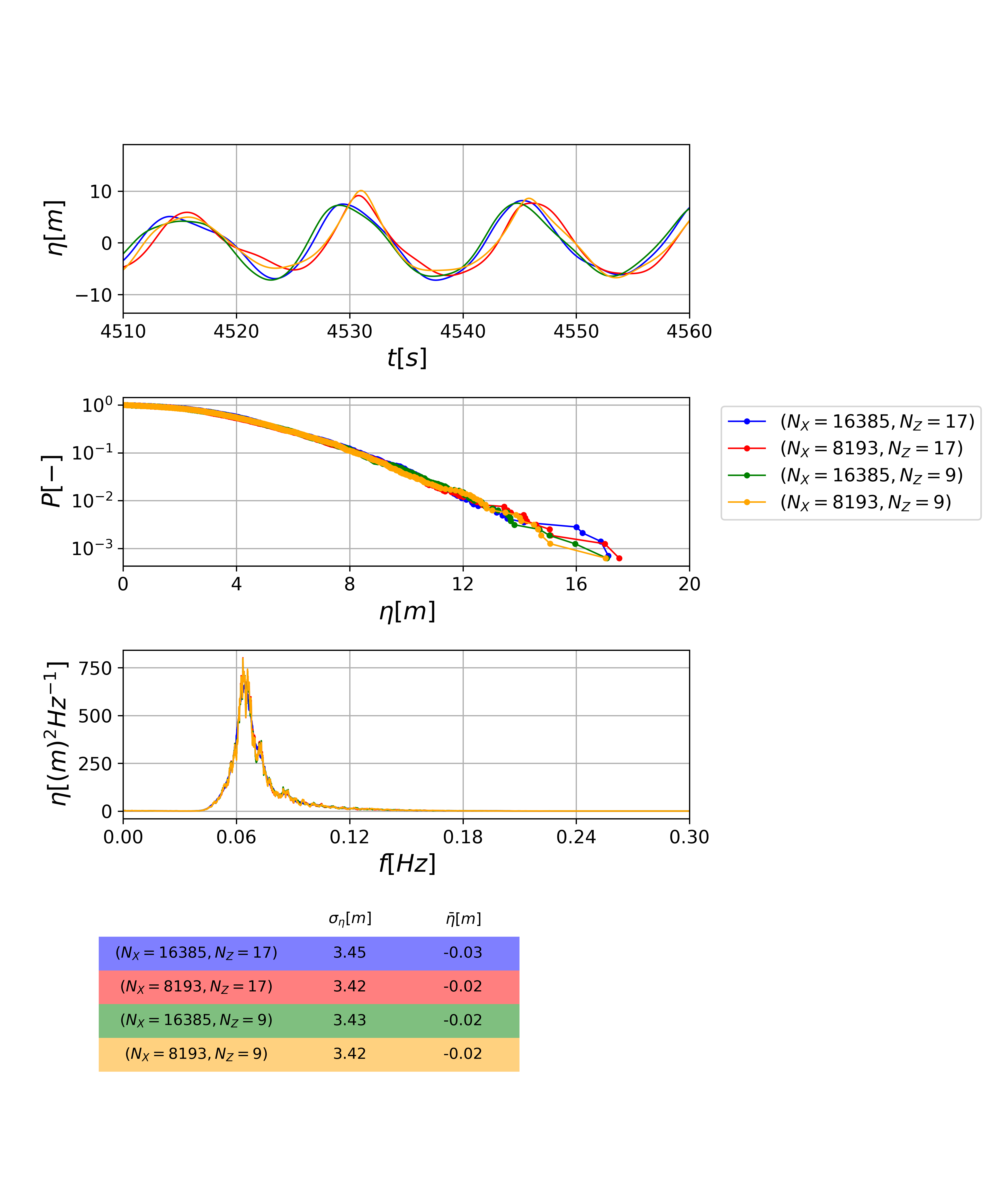}
  \caption{Grid dependency study. $H_S=15.77[m], T_P=15.15[s], \gamma=3.9$. Time series sampled at $h=75.0[m]$ and at $x=15\e 3[m]$.}
  \label{fig:grid_dep_deep}
\end{figure}

\begin{figure}[hbtp!]
  \centering
  \includegraphics[width=\textwidth, clip, trim={0 7.5cm 0 0}]{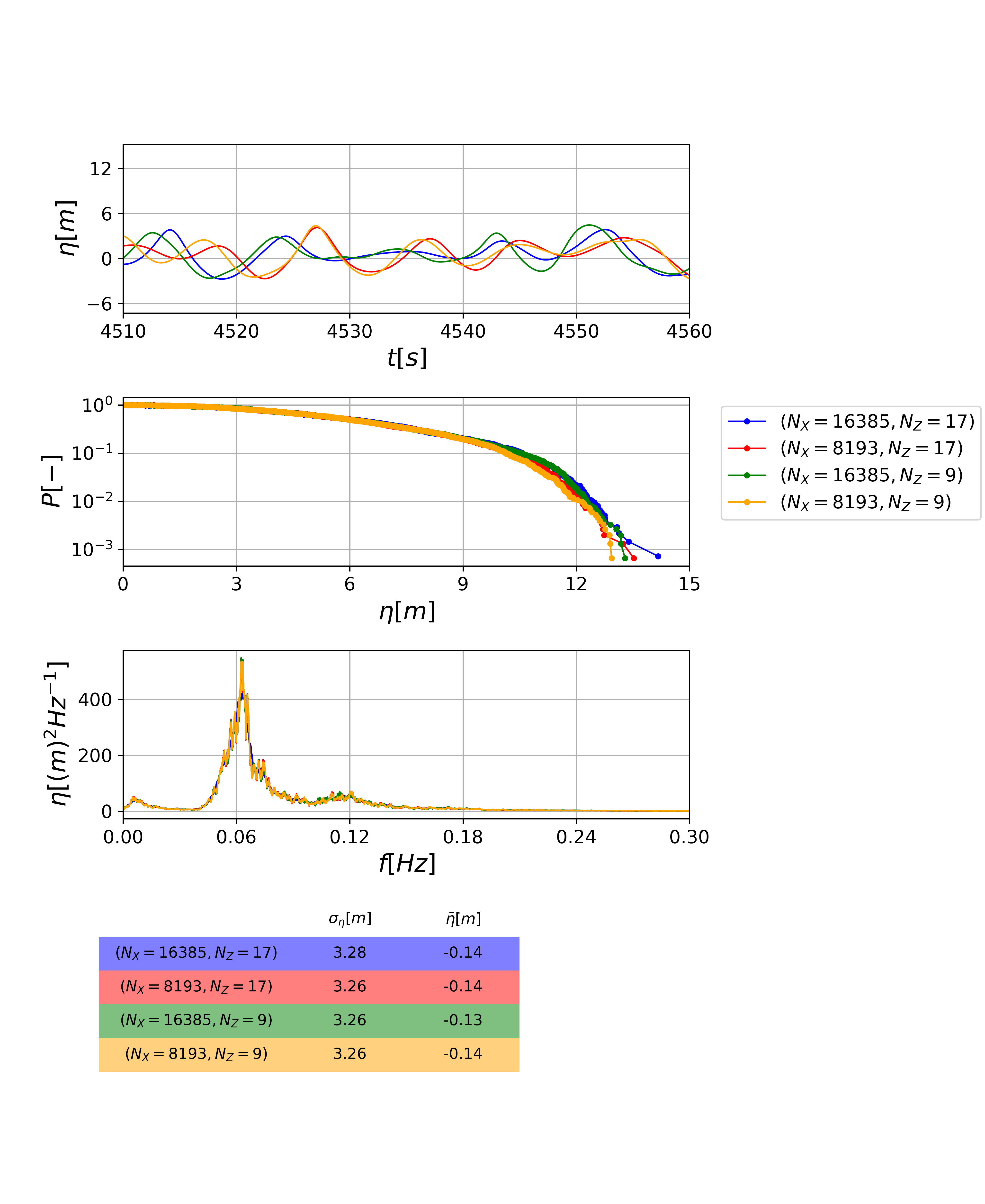}
  \caption{Grid dependency study. $H_S=15.77[m], T_P=15.15[s], \gamma=3.9$. Time series sampled at $h=25.0[m]$ and at $x=18.9\e 3[m]$.}
  \label{fig:grid_dep_shallow}
\end{figure}

\begin{figure}[hbpt!]
  \centering
  \includegraphics[width=0.8\textwidth]{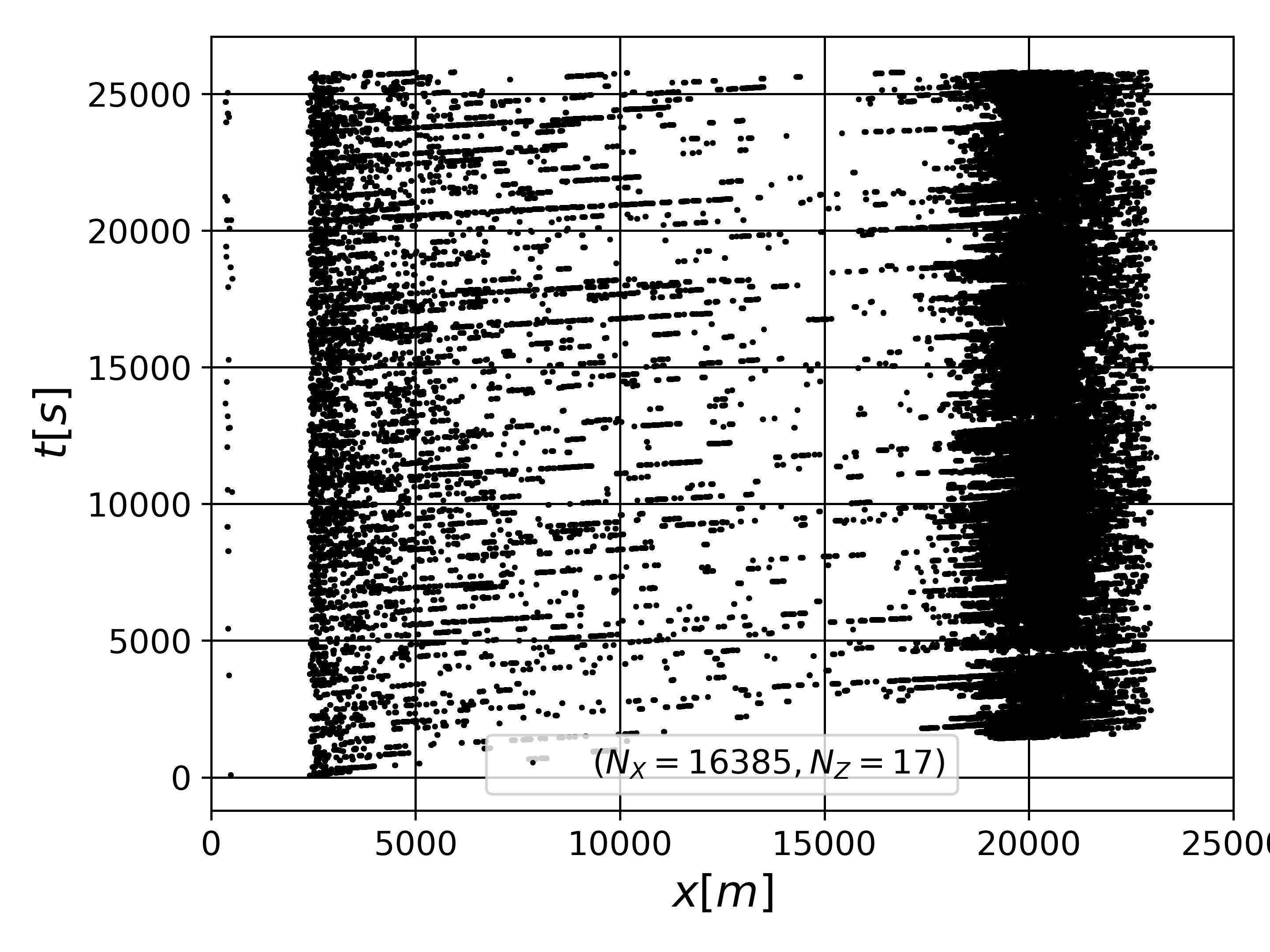}
  \caption{The black dots represent the locations and time instants at which the breaking filter ($\beta=0.5$) is active, for sea state 6 in \autoref{tab:database_sea_states}.}
  \label{fig:break_tracks}
\end{figure}

As a further check, we perform simulations with different grid discretizations on sea state 6 in \autoref{tab:database_sea_states}, keeping all the other parameters constant and using a breaking factor $\beta=0.5$.
In \autoref{fig:grid_dep_deep}, we can observe the effect of halving the grid resolution in the $x$-direction (red), in the $z$-direction (green), and in the $x$- and $z$-direction together (orange).
The time series was sampled at $x=15500[m]$, at a depth of $h=75.0[m]$, in a region that is characterized by a light breaking, see \autoref{fig:break_tracks}.
The power spectral densities are not affected by the change in discretization, while the peak statistics are. Halving the number of grid points either in the $x$- or in the $z$-direction brings along a $3 \%$ reduction in the crest heights for a probability level $P<3\e{-2}$, independent of the $x$-discretization. 
% However, the maximum crest height seems to be correctly captured. 

In the subset of the time signal plotted in the top subfigure, the discretization in the $x$ direction affects the phase of the waves. Curves obtained with the same $x$-wise discretizations seem to be in phase with each other (red and orange; green and blue). The $z$-wise discretization seems to impact on the sharpness of the peaks. Indeed, some phase error on the wavelengths above $kh=10.0$ is expected when $N_Z=9$ (see figure 8 of \citet{Engsig-Karup2009}), which can reduce or enhance local crest heights by changing the time at which wave components superimpose.
However, since the crest statistics are to a large extent unchanged, and since the chosen resolution is found here to be conservative, we conclude that reducing the discretization has a negligible impact on the amplitude errors of these small waves.

In \autoref{fig:grid_dep_shallow}, which was generated with data sampled at $x=18.9 \e 3[m]$ at a depth of $h=25.0[m]$, a similar tendency is visible. In this zone heavy breaking is occurring, cf. \autoref{fig:break_tracks}.
The top subplot confirms that the wave crests for signals with the same $x$-wise discretization are synchronized.
In the center subplot, a deviation between the two sets of simulations with $N_x=16385$ and $N_x=8193$ is visible for crests with low exceedance probability ($P<1 \e {-2}$). This suggests that reducing the $x$-wise discretization modifies the behavior of the breaking filter, which is overall more active and subtracts more energy. However changing the discretization has a negligible effect on the height of the largest crest. We conclude that our simulations with ($N_x=16385, N_z=17$) have converged to a satisfactory level.

% \todo{Simulations with $N_X=32769$ are running.}

\section{Application for load computation}
\label{sec:loadcomputation}
\subsection{Visualization of the Database}

\begin{figure}[hbtp]
  \centering
  \includegraphics[width=\textwidth]{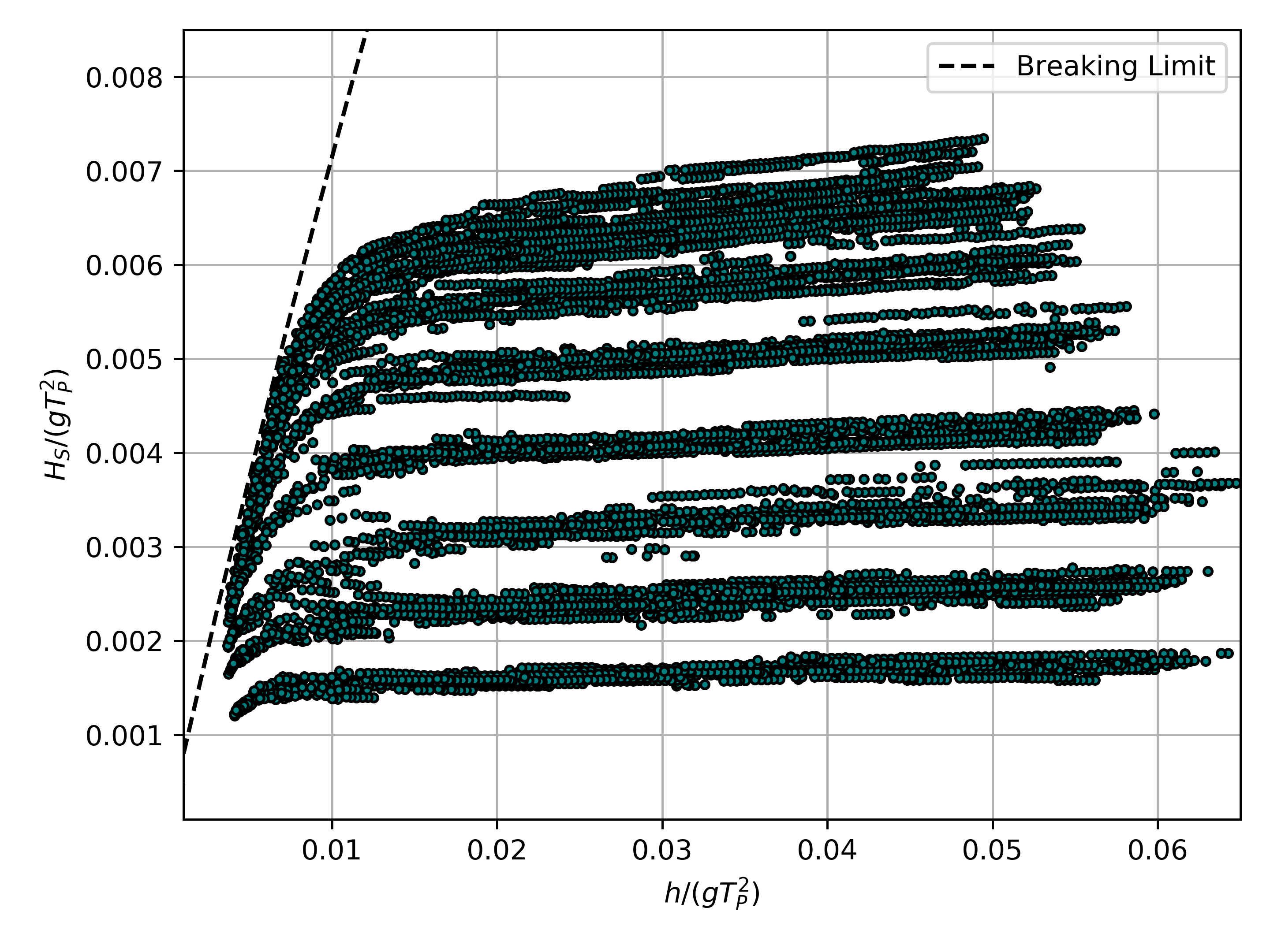}
  \caption{Realizations of the experiments collected in the database. The black dashed line is the theoretical breaking limit for regular waves.}
  \label{fig:database}
\end{figure}

In \autoref{fig:database}, all the 9648 sampled points are represented in terms of their $H_S^*$ and $h^*$.

\subsection{Extraction of kinematics and generalization via Froude scaling}

We denote as $(H_{S,0}, T_{P,0}, h_0)$ the combination of sea state and depth for which kinematics need to be produced. This tuple uniquely identifies a point in the $(h^*, H_{S}^*)$ plane. It is possible to extract from the database a number of realizations that are nondimensionally identical to the required sea state.
%as in \autoref{fig:database}, with a free surface elevation $\eta(x,y,t)$ and velocities and accelerations at a prescribed depth $u(x,y,z,t)$ and $u_t(x,y,z,t)$.

In the nondimensional plane $(h^*, H_{S}^*)$, the distance $\Delta$ of the $i^{th}$ sea state to the sea state of interest $(h^*_0, H_{S,0}^*)$ can be calculated as:

\begin{align}
  \Delta_i = \sqrt{(h^*_0-h^*_i)^2+ (H_{S,0}^{*}-H_{S,i}^{*})^2}
\end{align}

According to $\Delta_i$, we select the $N$ database points closest to the sea state of interest, where typically $N=5$. Only one realization per actual run of the solver is chosen since we are only interested in statistically independent simulations.

To match the scale of the sea state of interest, we need to compute a factor $S_i$ for each $i^{th}$ sea state, with $i=1..N$, by taking the ratio of the depth of the database point and the depth of the sea state of interest

\begin{align}
    S = \left(\frac{h_0}{h_i}\right)
    \label{eq:lambda}
\end{align}

Once the scaling factor is calculated, the simulated velocity, wave elevation, acceleration time series and time are Froude-scaled via

\begin{align}
  \eta_i^{\text{scaled}}(x,y,t) = \eta_i(x,y,t)S_i \\
  u_i^{\text{scaled}}(x,y,t) = u_i(x,y,t)S_i^{1/2} \\
  u_{t,i}^{\text{scaled}}(x,y,t) = u_{t,i}(x,y,t) \\
  t_i^{\text{scaled}} = t_i S_i^{1/2}
\end{align}

\subsubsection{Testing of the Froude scaling hypothesis}

We verify the appropriateness of the Froude scaling hypotheses by comparing the results of two simulations, one at full scale and one in a scaled domain.
The full scale simulation was performed on the domain in section \ref{sec:numerical_domain}. The second simulation was performed on a smaller domain, scaled by a geometrical factor $S=0.64$. The parameters for the scaled simulation are in \autoref{tab:froude_cases}.
The significant wave height and peak periods were also scaled to $H_S=10.09[m]$ and $T_P=12.12[s]$, while the peak enhancement factor was the same, $\gamma=3.9$. The scaled spectrum was cut at the same nondimensional frequency $kh_{{\text{in}}}=114.39$.
The number of grid points was also equal, with $(N_x,N_y,N_Z) = (16385, 1, 17)$, while the time step for the scaled simulation was reduced to $\Delta t=0.056 [s]$.

\begin{table}
  \centering
  \begin{tabular}{ccccccccc}
    \toprule
    -      & $H_S[m] $ & $T_P[s]$ & $\gamma$ & $L_x[m]$ & $h_{\text{in}}[m]$ & $h_{\text{out}}[m]$ & $h_{sample}[m]$ & $\Delta t[s]$ \\
    \midrule
    Original  & 15.77     & 15.15 & 3.9   & 25\e {3} & 250.0 &  12.5 & 25.0 & 0.07\\
    Scaled    & 10.09     & 12.12 & 3.9   & 16\e {3} & 160.0 &  8.0  & 16.0 & 0.056\\
    \bottomrule
  \end{tabular}
  \caption{The two sea states and domain configurations used to test the Froude Scaling Hypothesis.}
  \label{tab:froude_cases}
\end{table}

\begin{figure}[hbtp!]
  \centering
  \includegraphics[width=\textwidth, clip, trim={0 7.5cm 0 0}]{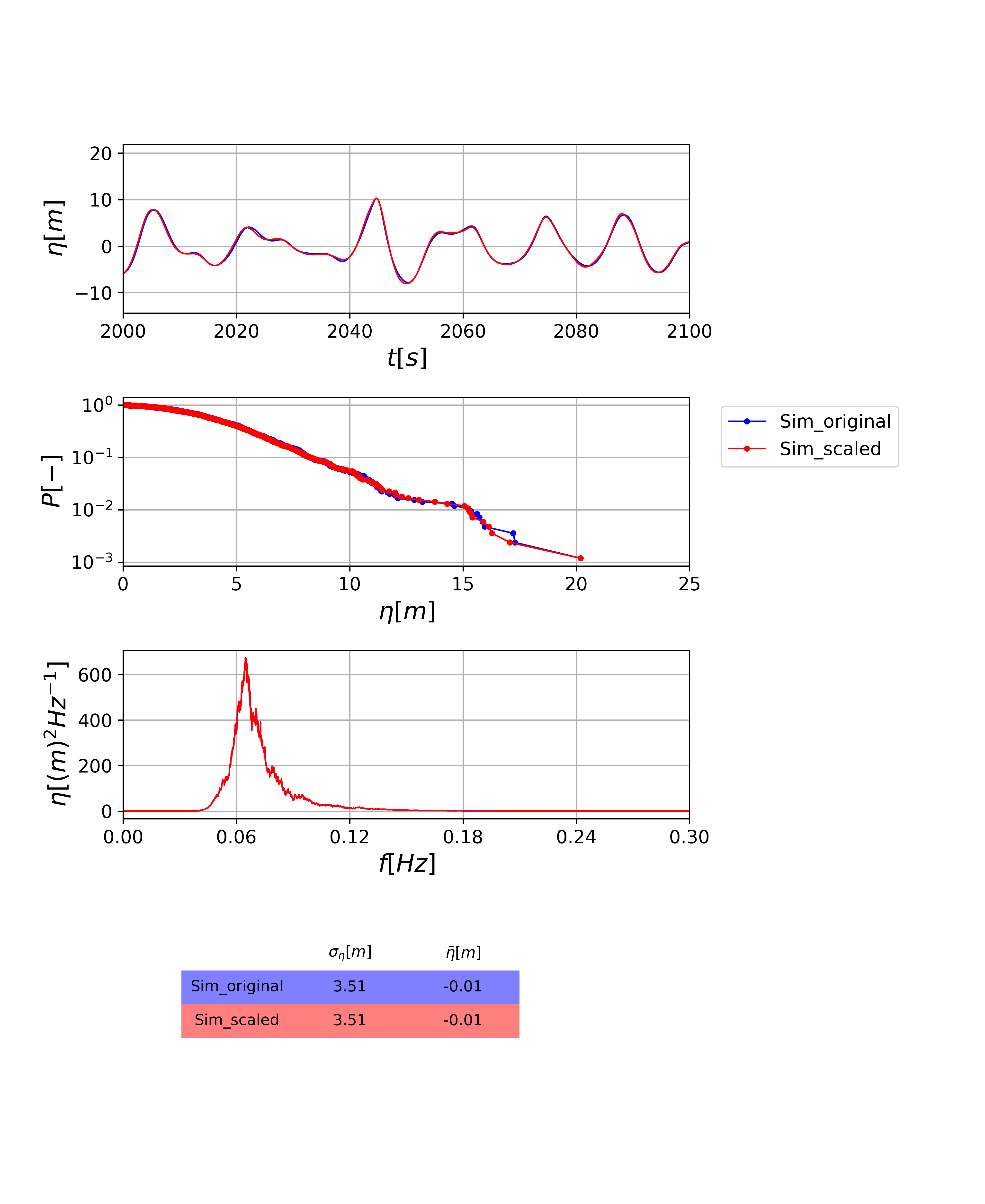}
  \caption{Comparison between the original and the scaled computation.}
  \label{fig:frouce_cases}
\end{figure}

In \autoref{fig:frouce_cases} we plot the statistics for a free surface elevation signal of 3 hours, sampled at $x=12.8 \e 3 [m]$, at a depth of $h=125.0[m]$. 
When the simulation in the small domain is scaled up, the peak exceedance probabilities of the free surface elevation are the same, with some minor discrepancies on the peak values. The same is true for the power spectral densities, which accurately match. Due to the successful deterministic comparison of the surface elevation, we infer that the kinematics behave accordingly. We conclude that the results sampled in a particular domain can be Froude scaled without loss of generality.

\subsection{The force models}

Once the wave kinematics are known, according to Rainey \cite{Rainey1995}, the horizontal sectional force $f_x$ on the cylinder is computed as

\begin{align}
    f_x = \frac{1}{2} \rho C_D D u|u| + \rho (1+C_m) \pi R^2 \frac{\text{d}u}{\text{d}t}
    \label{eq:rainey_dist}
\end{align}

where $\rho$ is the fluid density, $C_D$ is the drag coefficient, $C_m$ is the added mass coefficient, $R=D/2$ is the cylinder radius, and $\text{d}u/\text{d}t$ is the total (material) derivative of the streamwise velocity.

The horizontal axial divergence distributed force is computed by

\begin{align}
    f_{x,ad} = \rho \pi R^2 C_m w_z u
    \label{eq:rainey_ad}
\end{align}

where $w_z=\partial w / \partial z$. This force accounts for the fact that although the cylinder is slender in the horizontal direction, it is not slender in the vertical direction with respect to the wavelength.
The force at the monopile base is obtained by depth-integrating the distributed force contributions from the bottom of the monopile $z=-h$ to the time-varying surface elevation $z=\eta(t)$.
The distributed force from \autoref{eq:rainey_dist} and \ref{eq:rainey_ad} agrees with the recently published force model by \citet{KF_FNV}. 

A point force is added at the intersection between the surface and the cylinder, to take into account for the change of fluid kinetic energy which happens when the wet length of the cylinder varies

\begin{align}
    F_{\eta} = \frac{1}{2} \rho \pi R^2 C_m \eta_x u^2
    \label{eq:rainey_pf}
\end{align}

where $\eta_x={\partial \eta}/{\partial x}$. The Rainey load model is derived on energy conservation grounds only, and does not make any assumptions on the incoming waves, except for the fact that the cylinder is slender and does not modify the wave surface, also called the \textit{wavy lid assumption}.
While this point force differs from the $F^\psi$ in \citet{KF_FNV}, a recent study of \citet{Loup2020} concluded that the third-harmonic content of the two terms agree for large waves.

To find values for $C_M=C_m+1=2$ and $C_D$ for the smooth steel cylinder, a designer normally relies on recommended practices, as DNV-RP-C205 (see 6-7 and 6-9 in \cite{veritas2007recommended}). Generic values for a smooth cylinder at sub-critical Reynolds numbers are $C_m=1.0$ and $C_D=1.0$. We discuss our choice of parameters in the results section.

%%% Local Variables:
%%% mode: latex
%%% TeX-master: "../DeRiskDB"
%%% End:

%% file: texFiles/res.tex
\section{Comparison of Numerical and Experimental results}
\label{sec:results}

\subsection{The DeRisk experiments}

\begin{figure}[btp]
  \centering
  \includegraphics[width=0.75\textwidth]{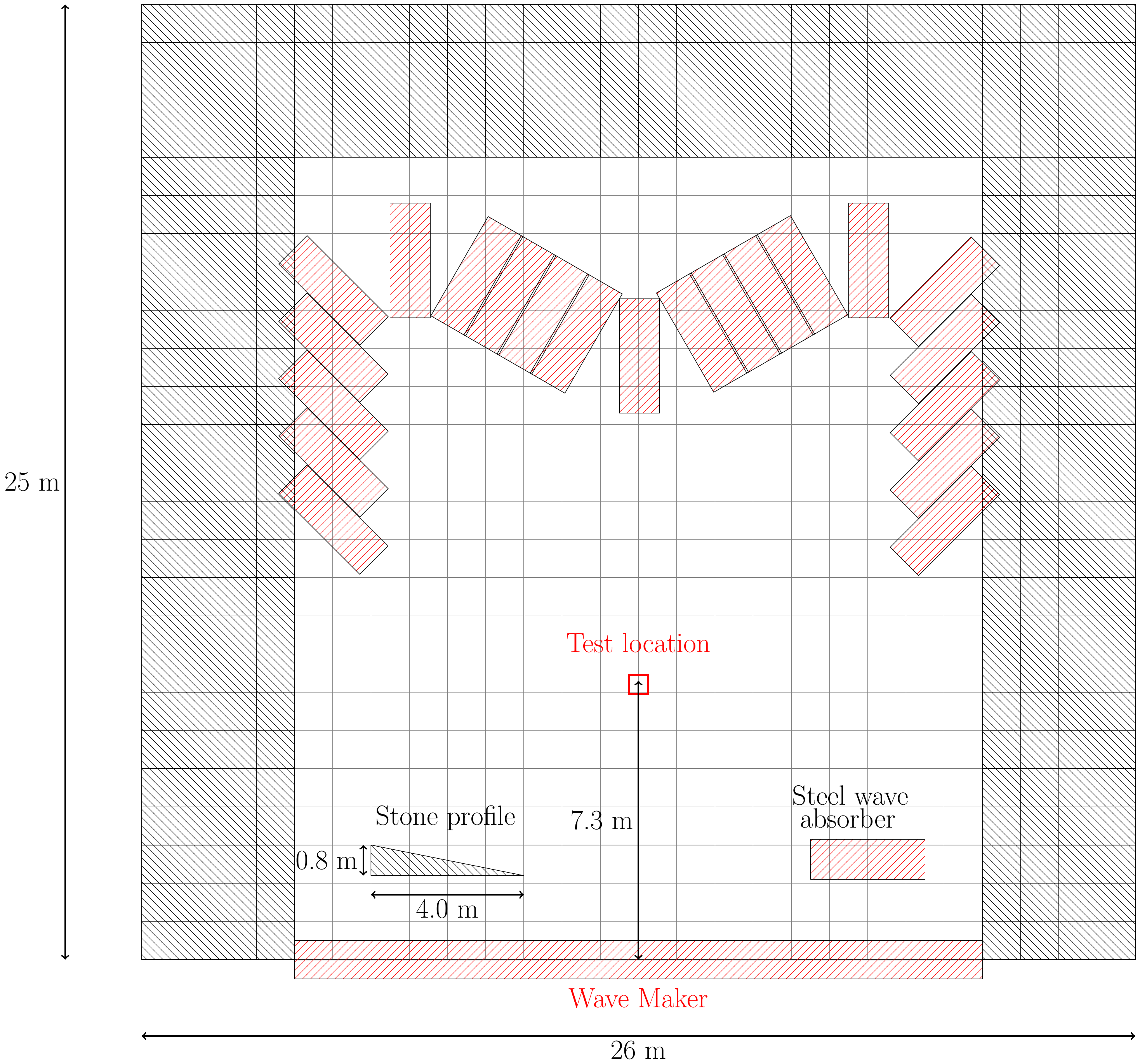}  
  \caption{DeRisk experimental setup in model scale, which is is 1:50 with respect to full scale. The grid resolution is $1[m] \times 1 [m]$. Adapted from Figure 1 in \citet{Bredmose2016}.}
  \label{fig:derisk_setup}
\end{figure}

The model tests used here for comparison were performed at DHI in 2015 in the framework of the DeRisk project.
The experiments aim to reproduce storms with different return times (10 to 1000 years) for two typical North Sea locations ($h=20[m]$ and $h=33[m]$ depth). 
Values in this section are in full scale, which is 50:1 to laboratory scale, except where otherwise indicated.
The experimental setup is sketched in \autoref{fig:derisk_setup}. 
% The wave tank was $W=1250[m]$ wide and $L=1500[m]$ long, with a flat bottom. 
The effective length of the basin is reduced by positioning arc-shaped steel wave absorbers in an M-shape, to direct the reflected waves away from the cylinder. To further mitigate wave reflection, crushed stones were arranged on the basin's edges with a slope of 1:5.

The PVC cylinder with a diameter of $D=7.0[m]$ is positioned $365.0[m]$ away from the paddles. 
The cylinder is supported by two force transducers, one at the top and one at the bottom, sampled at a frequency of $178.2[Hz]$.
The first bending eigenfrequency of the transducers and cylinder assembly is tuned to be far away from the wave spectrum frequencies to reduce the dynamic structural response.

The basin's wave maker is made up of 36 piston-type wave paddles and is capable of generating both 2D and 3D waves. The wave paddle signal for each of the test runs is stored, so that it can be reproduced numerically at a later stage.
An array of $5\times5$ wave gauges is positioned around the cylinder with a full-scale spatial resolution of $10[m]$. Five additional wave gauges are located in the space between the wave gauge array and the wave paddle generator. The wave gauge signal is sampled at $12.7 [Hz]$.
A 3D Vectrino velocimeter is positioned in the wave tank right in front of the cylinder, located at half the wave tank's water depth.

In this work, all the experiments are analyzed, but we only present plots for experiments 11 ($h=33.0[m]$) and 23 ($h=20.0[m]$). In these tests, two steep sea states with a 100-year (yr) return period were applied, and we expect the waves to be strongly nonlinear.

\begin{table}[btp]
  \centering
  \small
  \begin{tabular}{c|ccc|cc|cc}
    \toprule
       & \multicolumn{3}{c}{Nominal}  & \multicolumn{2}{c}{Measured} & \multicolumn{2}{c}{}\\
    \midrule
    No. &  $H_S$  & $T_P$ & $\gamma$ & $H_S$ & $T_P$ & Depth[m] & Return [yr] \\
    \midrule
    \input{tables/derisk_exp.tex}
    \bottomrule
  \end{tabular}
  \caption{Experimental tests with long-crested waves. We report both the nominal values, imposed at the boundary and the measured values at the cylinder location.
    }
  \label{tab:derisk_exp}
\end{table}
 
\subsection{Reproduction of experiments on a flat domain with boundary-flux wave generation}
\label{sec:paddle}

To validate the flow model, we replicate the DeRisk experimental results via OceanWave3D.
In a previous study by \citet{schloer2017experimental}, experiments 23 and 24 of \autoref{tab:derisk_exp} are reproduced deterministically via OceanWave3D by using the experimental wave paddle signal to drive the flux at the numerical wave generation boundary. The numerical domain has a full-scale size of $L_x=1000.[m]$, a flat bottom with a constant depth of $h=20.0[m]$, and was discretized into $(N_x,N_y,N_Z) = (201, 1, 15)$ grid points. The distance between the sampling point and the wave generator is $365.0[m]$, as in the experimental domain (see \autoref{fig:derisk_setup}).
Part of the discrepancy between the measured and the target significant wave height is due to breaking. For the 1000-yr storms, breaking was visually observed right in front of the paddle. However, it is difficult to quantify the amount of energy lost due to the viscous dissipation process.

To calibrate our simulations, we reproduce Experiments 11 and 23 using the same computational domain and grid discretization as in \citeauthor{schloer2017experimental}, and we compare numerical and experimental free surface elevation and Eulerian acceleration.
The results in \autoref{fig:Test11}, relative to the 100-yr storm at depth $h=33.0[m]$ obtained with $\beta=0.5$, show a good agreement. The measured energy distribution at higher frequencies ($f>0.1[Hz]$) is also correctly reproduced for the horizontal Eulerian acceleration, when compared to measurements by the Vectrino sensor positioned at $z=-16.5[m]$, despite what seems like a small ($\approx 0.1[m/s]$) magnitude systematic error in the exceedance probability.

\begin{figure}[btph!]
  \centering
  \begin{subfigure}{0.9\textwidth}  
    \includegraphics[width=0.9\linewidth]{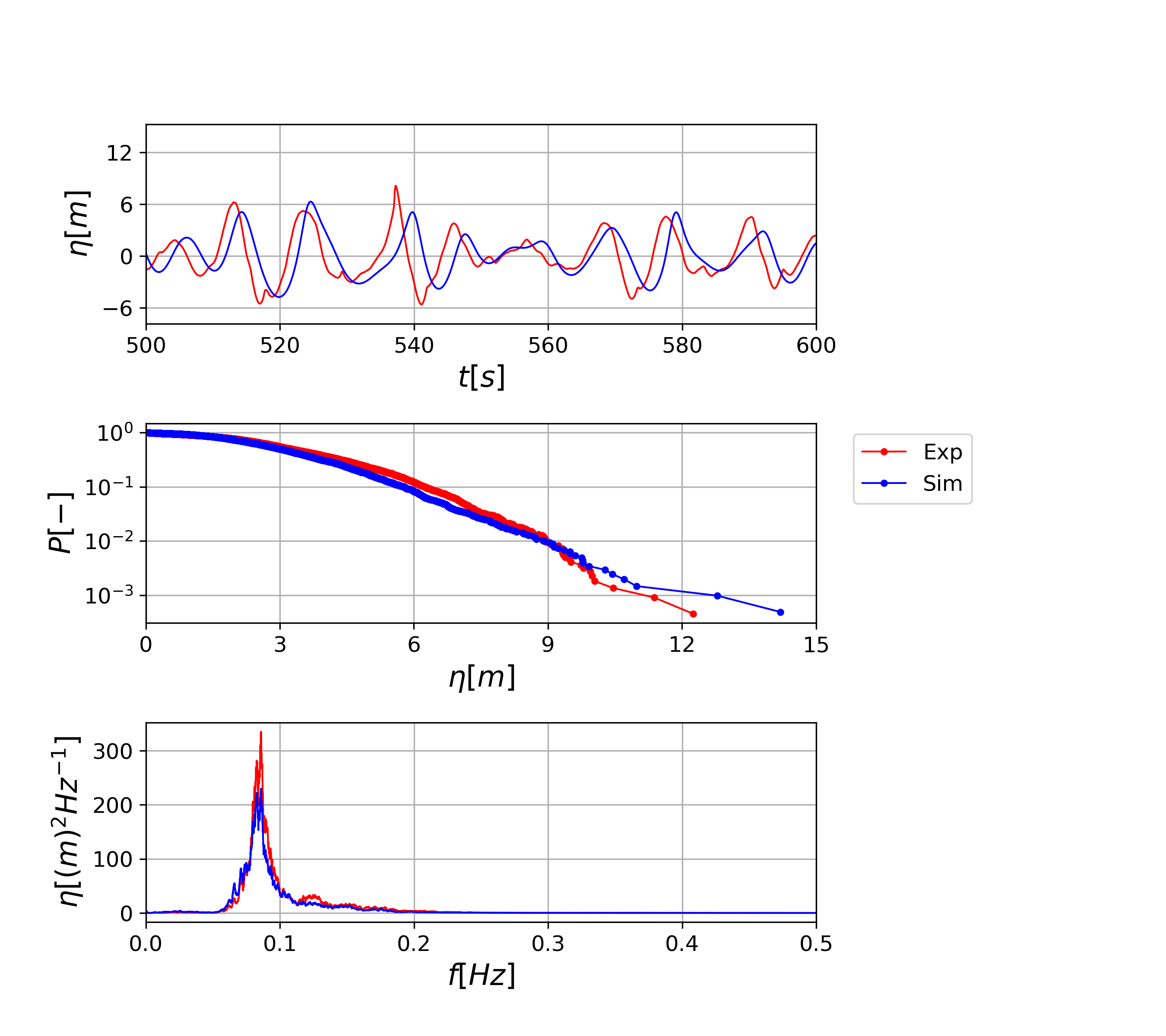}
    \caption{Plot of 3 hours of the OceanWave3D wave elevation signal, from simulations and experiment 11 in \autoref{tab:derisk_exp}, $\beta=0.5$. Good agreement in the power spectrum and in the crest exceedance probability.}
    \label{fig:sim_exp_11}
  \end{subfigure}
\end{figure}
\begin{figure}[h!]
  \centering
  \ContinuedFloat
  \begin{subfigure}{0.9\textwidth}
    \includegraphics[width=0.9\textwidth]{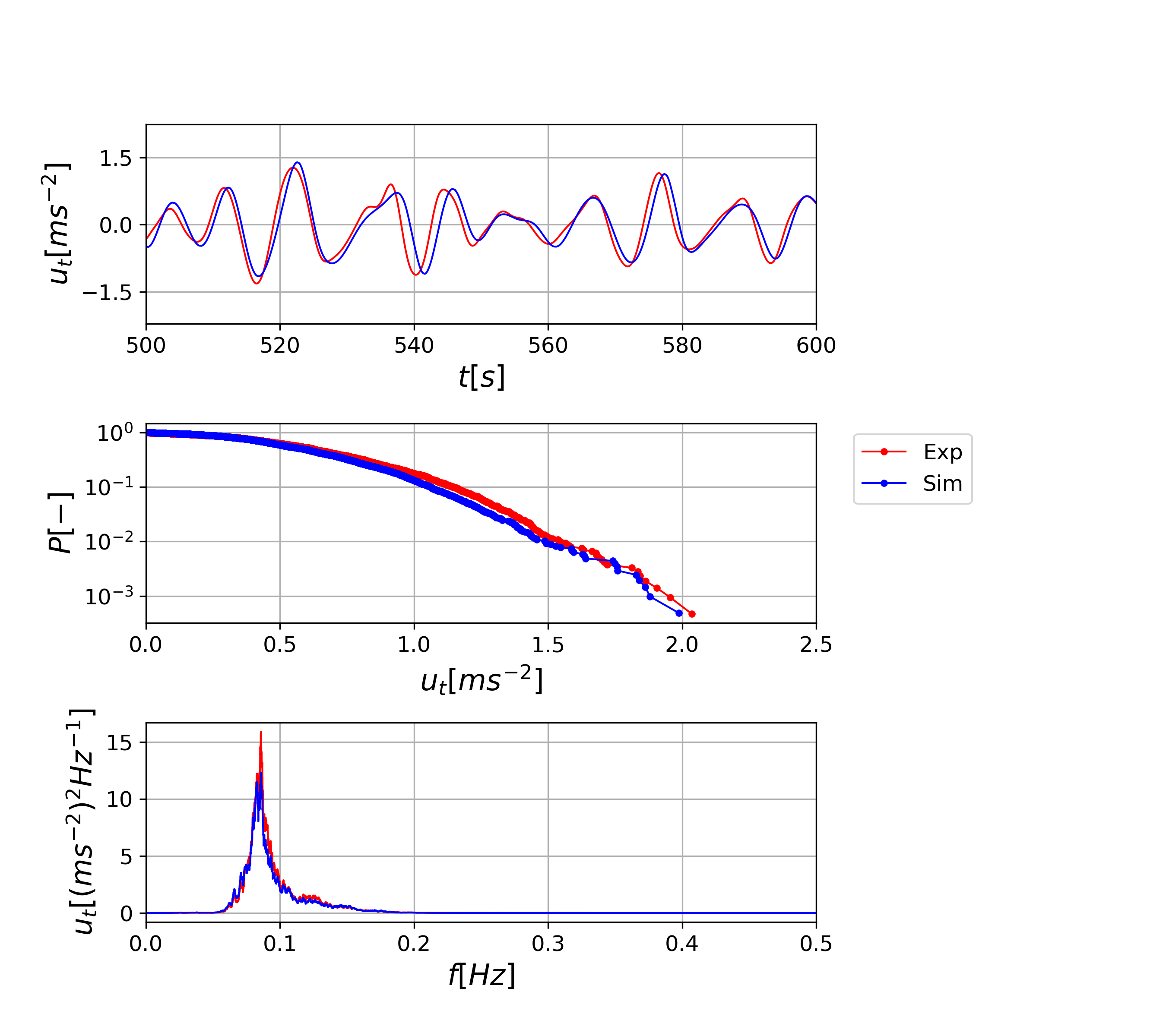}
    \caption{Plot of 3 hours of the OceanWave3D wave horizontal acceleration $u_t$ at $z=-16.5[m]$, from simulations and experiments 11 in \autoref{tab:derisk_exp}, $\beta=0.5$.}
    \label{fig:ut_sim_exp_11}  
  \end{subfigure}
\caption{Test 11 reproduced with the paddle signal.}
\label{fig:Test11}
\end{figure}

For the 100-yr storm in the $h=20.0[m]$ depth are in \autoref{fig:Test23}, the best agreement between the wave elevation measurements and the simulated free surface elevation power spectral density (PSD) was obtained with $\beta=0.3$. In particular, the superharmonic peak location and magnitude are very well reproduced, even though some energy is missing in the main peak. The same comparison is in Figure 6 of \citeauthor{schloer2017experimental}, which was produced by data sampled on an equivalent computational domain.
The peak exceedance probability agrees well down to $P=2.5\e{-2}$ but diverge for larger values. We regard this as an underestimation of breaking by the simulations.

In \autoref{fig:u_sim_exp_23}, the computed and measured horizontal Eulerian acceleration is compared at $z=-10.0[m]$. The acceleration spectrum is very well estimated by the simulation, even though some energy is missing around the peak frequency. 
The agreement of the exceedance probability is overall good, although a small deviation is observed for points with probability levels of $P< 1.5\e{-2}$.

\begin{figure}[btph]
  \centering
  \begin{subfigure}{0.9\linewidth}  
    \includegraphics[width=0.9\linewidth]{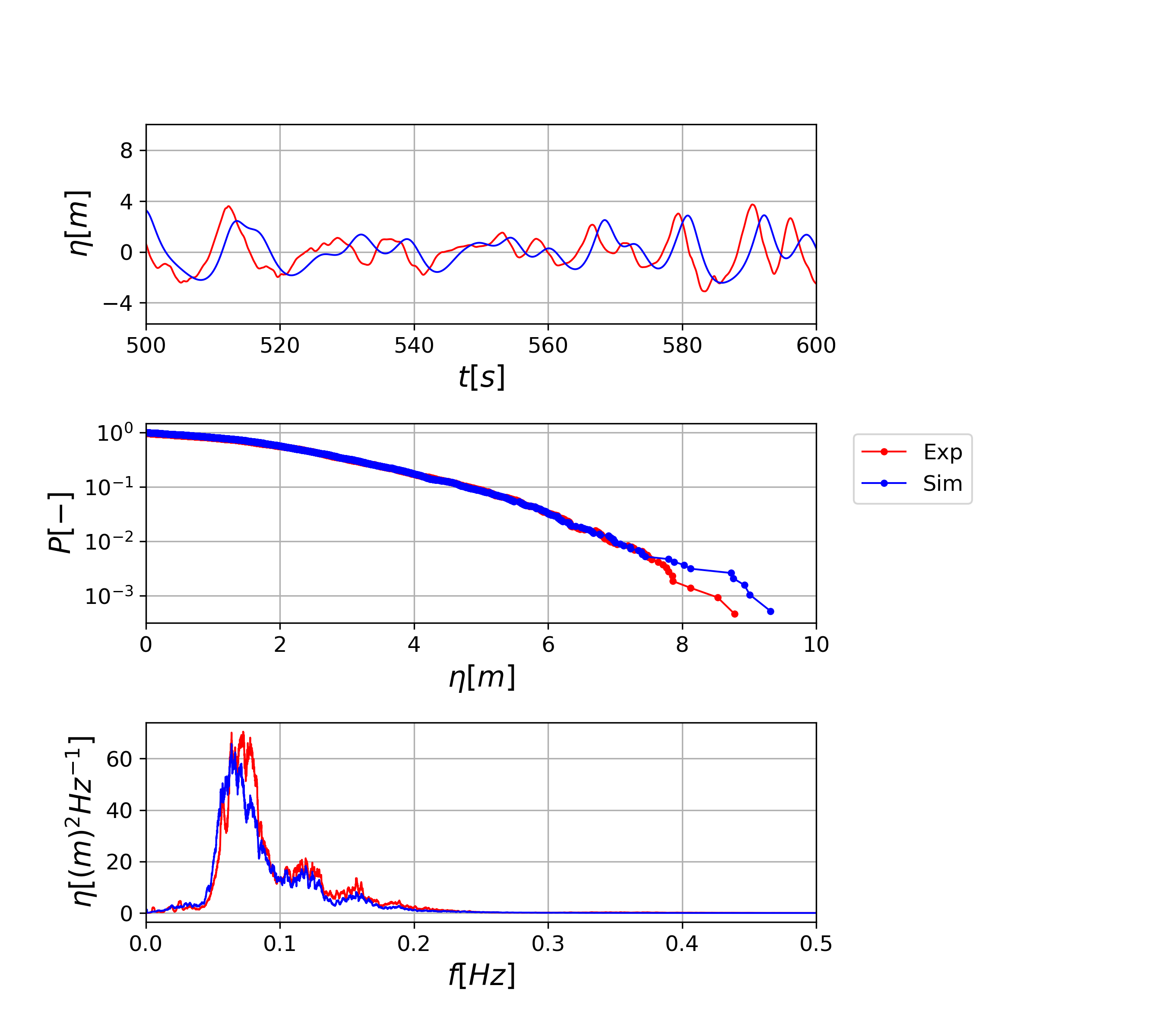}
    \caption{Plot of 6 hours of the OceanWave3D wave elevation signal, from simulations and experiment 23 in \autoref{tab:derisk_exp}, $\beta=0.3$. The spectra show a good agreement, except close to the spectral peak. The free surface elevation crests statistics deviate for larger waves, probably due to the effect of the breaking filter.}
    \label{fig:sim_exp_23}
  \end{subfigure}
\end{figure}
\begin{figure}[h!]
  \ContinuedFloat
  \begin{subfigure}{0.9\linewidth}
    \includegraphics[width=0.9\textwidth]{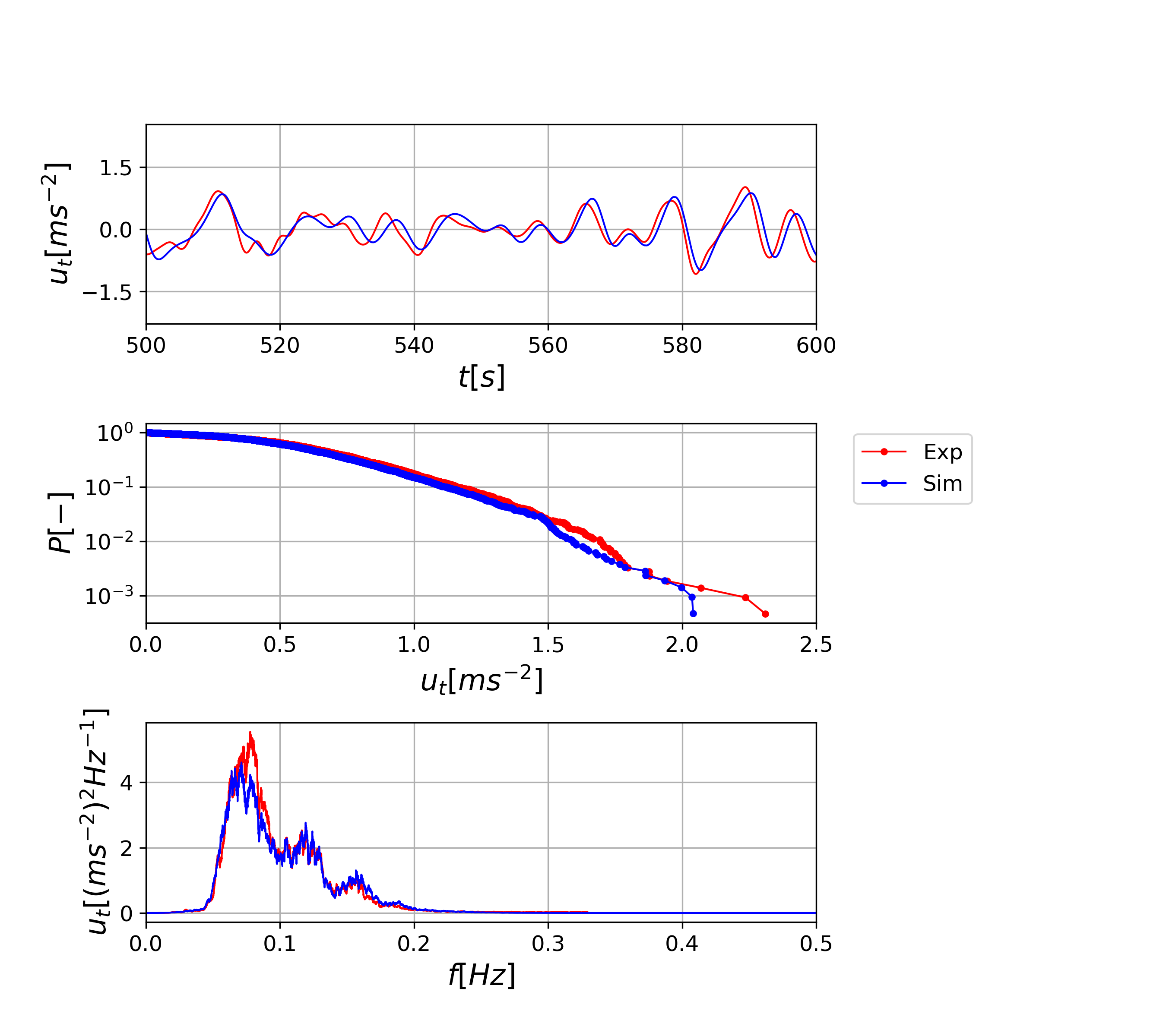}
    \caption{Plot of 6 hours of the OceanWave3D wave horizontal kinematics acceleration $u_t$ at $z=-10.0[m]$, from simulations and experiment 23 in \autoref{tab:derisk_exp}, $\beta=0.3$. Good agreement in the spectrum distribution and in the crest exceedance probability.}
    \label{fig:u_sim_exp_23}  
  \end{subfigure}
\caption{Test 23 reproduced with the paddle signal. Good agreement both for velocity and acceleration.}
\label{fig:Test23}
\end{figure}

We conclude that the potential flow solver with the flux wave generation reproduces the measurements fairly well, although this approach is quite sensitive to the $\beta$ factor.

\subsection{Reproduction of experiments on a flat domain with waves in a relaxation zone: effect of the wave breaking factor $\beta$}

% \todo{Try to take only the closest point. Question is: why is the force overestimated anyway? Even when the second order peaks are ok? Does this happen for Amin and Signe's paper as well? Answer: there is an error in the indexing of the kinematics in Signe's paper.}

We aim to calibrate the optimal breaking factor $\beta$ for the DeRisk database simulations, where the waves are generated in a relaxation zone rather than with a boundary flux.
We thus modify the flat-bed computations in section \ref{sec:paddle} to generate waves by random realizations in a relaxation zone rather than by imposing the boundary flux.
% by using smaller, faster computations.
% Since the phases are randomly chosen, we are only able to perform 
% a statistical comparison.
% It is less sensitive to the $\beta$ factor since it allows to do a statistical comparison rather than a deterministic one. 
% In this case, the small flat domain allows us to obtain results faster.
The full-scale domain length in these simulations is $L_x=1350.0[m]$, where the first $350.0[m]$ are used as the buffer zone for wave generation.
The domain was discretized into $(N_x,N_y,N_Z) = (513, 1, 17)$ grid points.
For the two combinations of $H_S$ and $T_P$ of cases 11 and 23 from \autoref{tab:derisk_exp}, 10 realizations were produced by using 10 different sets of random phases. Each realization was then repeated with three different breaking filter constants, $\beta=[0.3, 0.4, 0.5]$. 
% Our starting value for the breaking filter tolerance is $\beta=0.5$.

As for Test 11, at $h=33.0[m]$, in \autoref{fig:etaFlat11_gbCompare}, we notice a very good match between the measured and the simulated wave elevation signal statistics for $\beta=0.5$. A similar agreement was found in \autoref{fig:sim_exp_11}, even though the power spectrum of the boundary flux generated waves showed a better agreement in the super-harmonic region. 
The breaking factor does not significantly influence the shape of the spectrum, while the surface elevation peak exceedance probability curve shows a marked reduction for $P<1\e{-1}$ when $\beta=0.3$.

The statistics of the horizontal Eulerian acceleration, in \autoref{fig:utFlat11_gbCompare}, agree well for $\beta=0.5$. The exceedance probability for the kinematic acceleration peaks, though, diverged for probabilities lower than $P=2\e{-2}$, in contrast to \autoref{fig:ut_sim_exp_11} where the tail was well captured. In the case of the statistical comparison, we experience some variability in the tail of the distribution, determined by randomness.
The peak exceedance probability for the Eulerian acceleration was less sensitive to $\beta$ than for the free surface elevation, as the high-frequency components that are mostly affected by the smoothing filter do not penetrate down to $z=-16.5[m]$. For test 11, $\beta=0.5$ gave the best agreement overall.

In test 23, in \autoref{fig:etaFlat23_gbCompare}, the experimental and numerical secondary peaks in the super-harmonic region of the PSD match both in location and magnitude for $\beta=0.5$. However, the computations seem to underestimate the magnitude of the main spectral peak, as in \autoref{fig:sim_exp_23}.
The exceedance probability for $\beta=0.3$ underestimates the crest height, in contrast to \autoref{fig:sim_exp_23}. This suggests that the wave generation in a relaxation zone led to an overall larger amount of breaking, providing a better agreement for a less aggressive breaking constant of $\beta=0.5$.

The statistics of the streamwise Eulerian acceleration for $\beta=0.5$ agree well with the experimental ones (see \autoref{fig:utFlat23_gbCompare}).
The main spectral peak of the PSD is weaker than the experimental one and the secondary peak at $f=0.16[Hz]$ is overestimated.
These super-harmonic peaks are due to free waves generated at the inlet boundary and are therefore sensitive to the wave generator type. This aspect will be the subject of a follow-up paper.
We expect a less accurate agreement when the wave generation is distributed in a buffer zone rather than concentrated into a boundary flux.
The computed peak exceedance probability reproduced the experiments with an acceptable accuracy down to $P=3\e{-3}$, despite a small positive bias for $\beta=0.5$.
% When a more aggressive breaking filter is used, the magnitude of the secondary spectral peak at $f=0.16[Hz]$ was significantly affected. This suggests that the high-frequency components penetrate down to $z=-10.0[m]$.
We conclude that $\beta=0.5$ again returns the best agreement with the experiments.

\begin{figure}[htbp]
  \centering
  \begin{subfigure}[b]{0.9\linewidth}
    \includegraphics[width=0.95\linewidth]{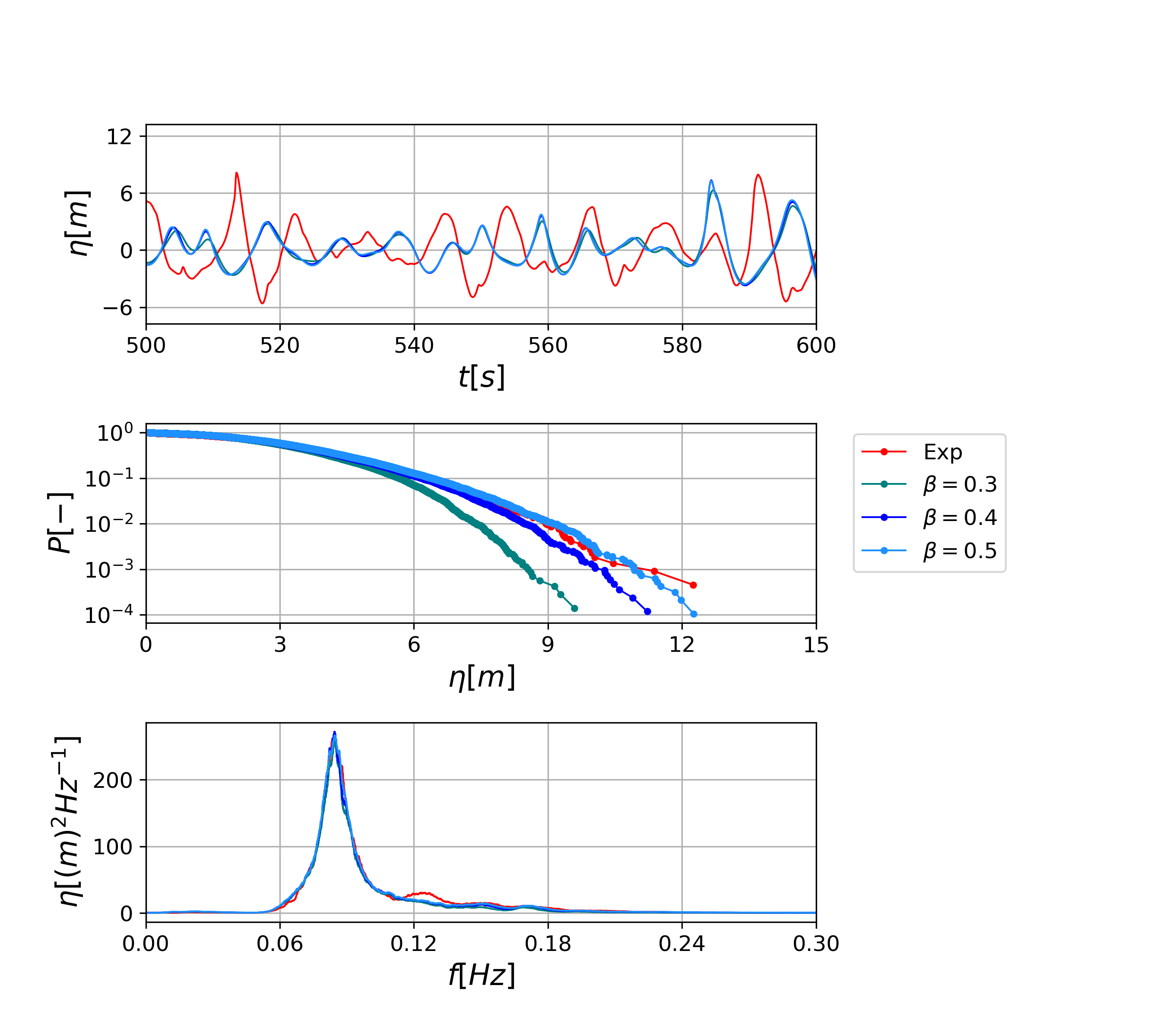}
    \caption{Case 11, 100-yr storm, $h=33.0[m]$, with $\beta=[0.3, 0.4, 0.5]$. Surface elevation.}
        \label{fig:etaFlat11_gbCompare}  
  \end{subfigure}
\end{figure}
\begin{figure}[h!]
  \ContinuedFloat
  \begin{subfigure}[b]{0.9\linewidth}
    \includegraphics[width=0.95\linewidth]{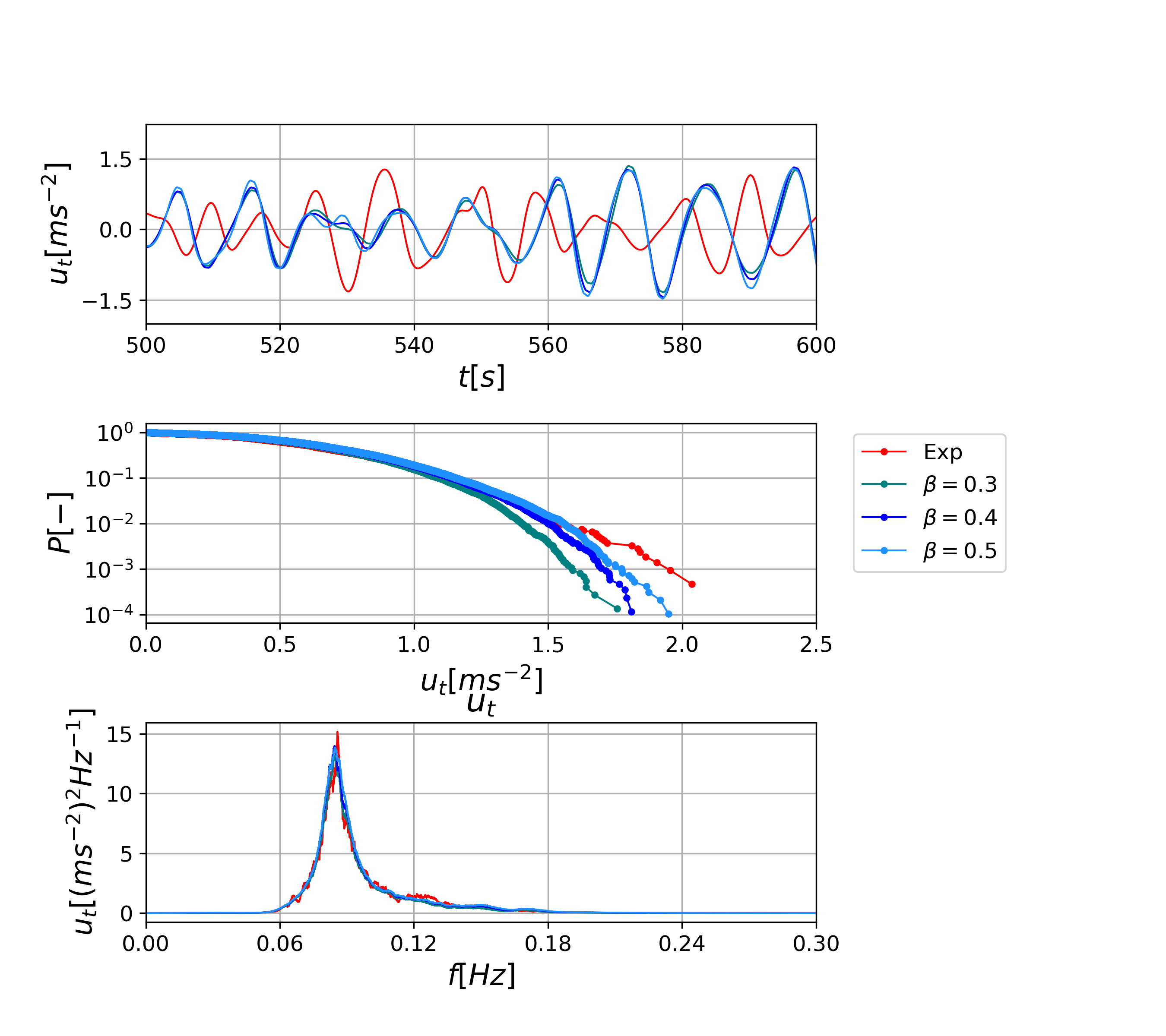}
    \caption{Case 11, 100-yr storm, $h=33.0[m]$, with $\beta=[0.3, 0.4, 0.5]$. Horizontal Eulerian acceleration at $z=-h/2$.}
        \label{fig:utFlat11_gbCompare}  
  \end{subfigure}  
  \caption{Case 11: comparison with the flatbed database.}
\end{figure}

\begin{figure}[htbp]
    \centering
    \begin{subfigure}[b]{0.9\textwidth}
      \includegraphics[width=0.95\textwidth]{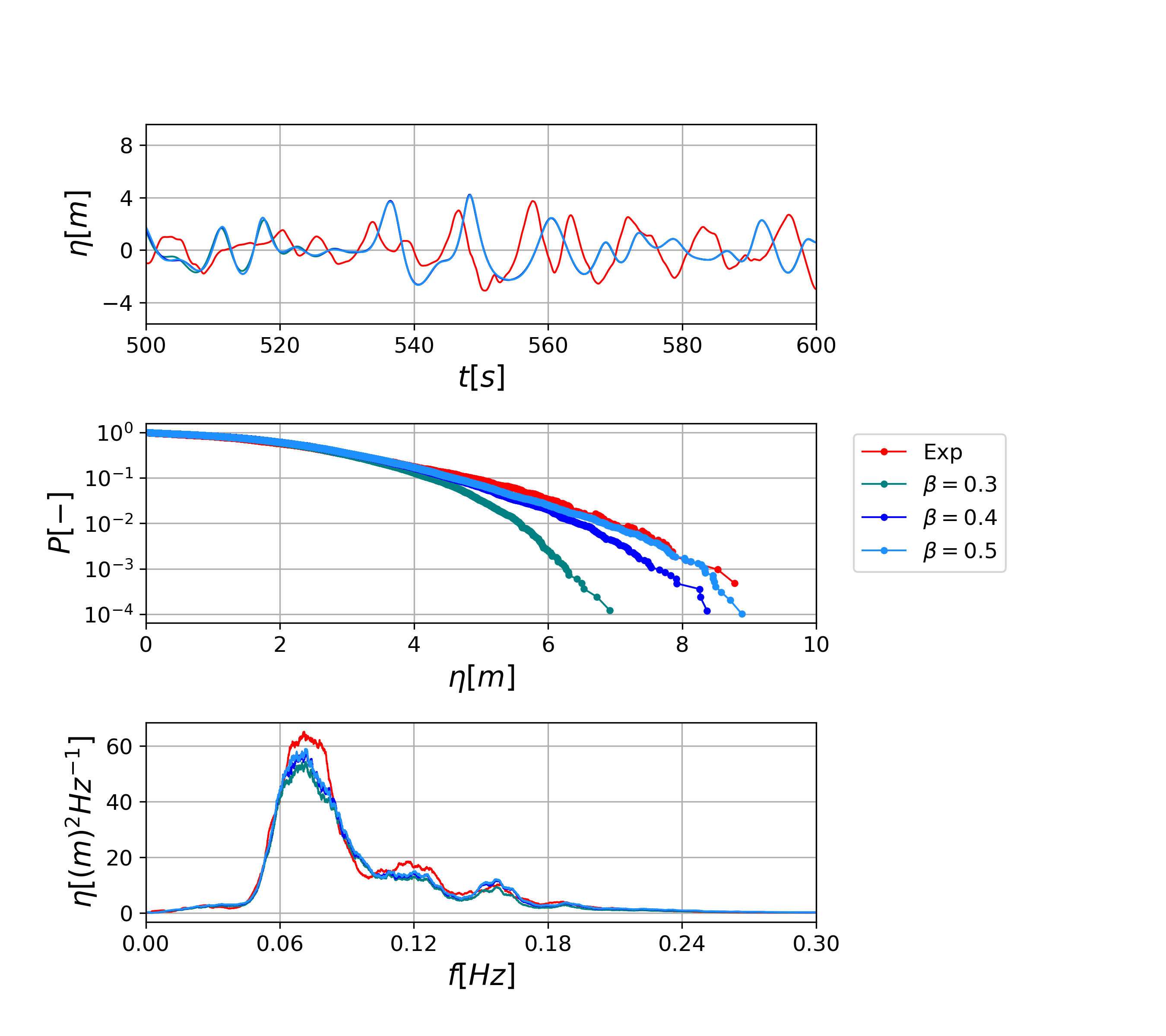}
      \caption{Case 23, 100-yr storm $h=20.0[m]$, with $\beta=[0.3, 0.4, 0.5]$. Surface elevation.}
          \label{fig:etaFlat23_gbCompare}  
    \end{subfigure}
  \end{figure}
  \begin{figure}[h!]
    \ContinuedFloat
    \begin{subfigure}[b]{0.9\textwidth}
      \includegraphics[width=0.95\textwidth]{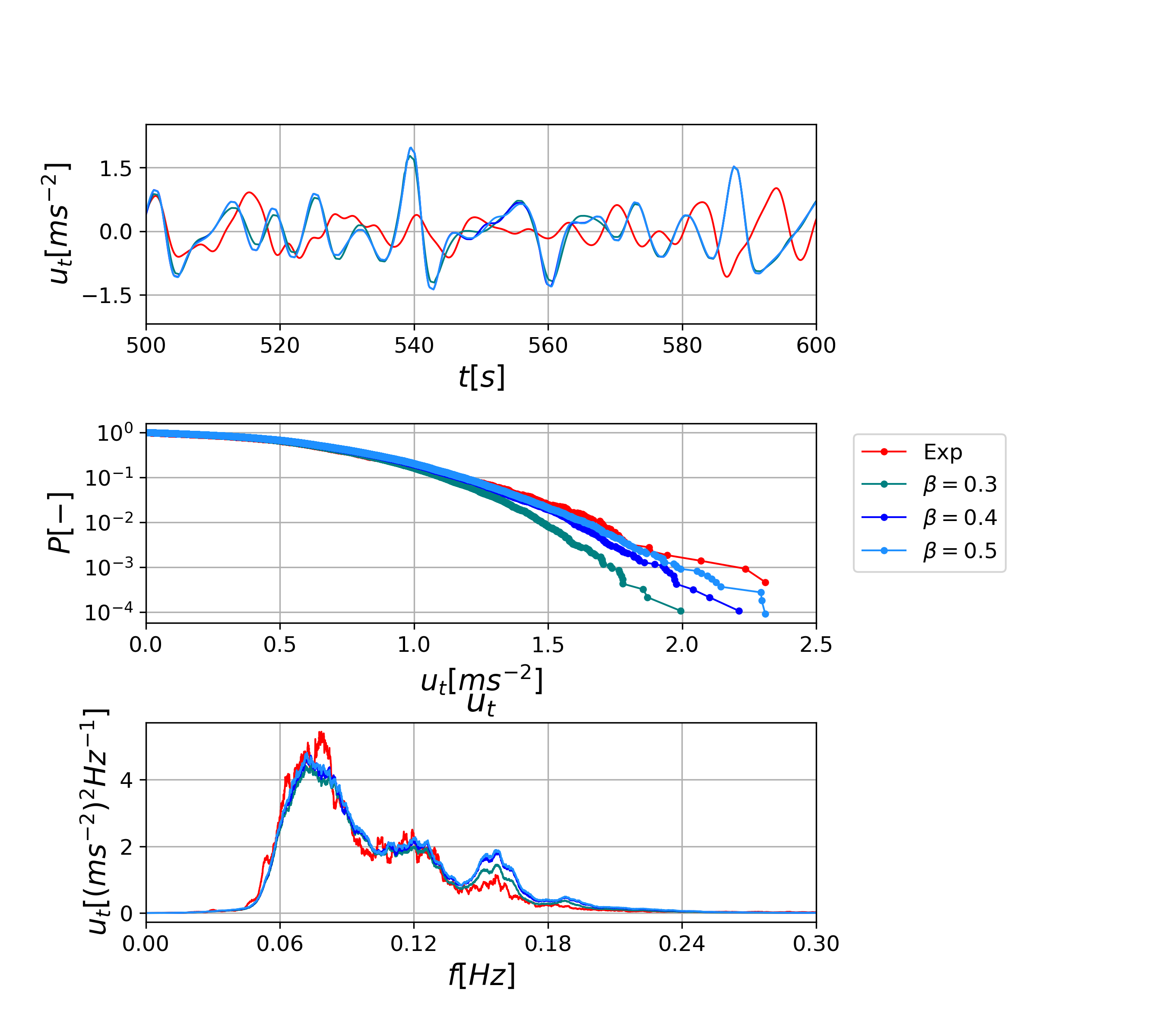}
      \caption{Case 23, 100-yr storm, $h=20.0[m]$, with $\beta=[0.3, 0.4, 0.5]$. Horizontal Eulerian acceleration at $z=-h/2$.}
          \label{fig:utFlat23_gbCompare}  
    \end{subfigure}    
    \caption{Case 23: comparison with the flat-bed database.}
\end{figure}

The value $\beta=0.5$ is not in agreement with the suggestions in \citet{schloer2017experimental}, which suggested $\beta=0.3$ for unidirectional waves. We believe this difference is justified by the different wave generation technique.
In the \autoref{tab:gBreak}, we collect the optimal $\beta$ values for all the analyzed cases. We find that $\beta=0.5$ works well across the full range of the investigated $h/gT_P^2$ and $H_S/gT_P^2$.

\begin{table}[h!]
  \centering
  \begin{tabular}{cccccc}
    \toprule
    Test N. & $h/gT_P^2$ &  $H_S/gT_P^2$  & $\beta$ & $C_M^*$ & $KC$\\
    \midrule
    1 &2.23\e{-2} &6.17\e{-3} & 0.5 & 1.70  & 2.89\\ 
    9 &2.70\e{-2} &6.37\e{-3} & 0.5 & 1.72  & 2.44 \\ 
    10 &1.82\e{-2} &4.05\e{-3} & 0.5 & 1.71  & 2.37 \\ 
    11 &2.49\e{-2} &7.41\e{-3} & 0.5 & 1.74  & 2.93 \\ 
    12 &1.82\e{-2} &5.05\e{-3} & 0.5 & 1.71  & 2.90 \\ 
    13 &1.80\e{-2} &5.59\e{-3} & 0.5 & 1.84  & 3.56 \\
    20 &1.39\e{-2} &4.21\e{-3} & 0.5 & 1.75  & 2.21 \\ 
    21 &1.32\e{-2} &4.08\e{-3} & 0.5 & 1.73  & 2.31 \\ 
    22 &1.36\e{-2} &4.82\e{-3} & 0.5 & 1.75  & 2.53 \\ 
    23 &1.03\e{-2} &3.63\e{-3} & 0.5 &  1.64 & 2.89 \\ 
    24 &1.03\e{-2} &3.95\e{-3} & 0.5 &  1.70 & 3.04 \\ 
    25 &2.62\e{-2} &7.98\e{-3} & 0.5 &  1.66 & 1.73 \\         
    \midrule
    $\bar{\square}$ & & & & 1.72 & \\
    $\sigma_\square$ & & & & 0.048 & \\
    \bottomrule 
  \end{tabular}
  \caption{Optimal breaking factors. In general, $\beta=0.5$ was found to give the best agreement. KC was based on the standard deviation of the streamwise surface velocity $KC=\sigma_{u|_{z=\eta}} D T^{-1}$.}
  \label{tab:gBreak}
\end{table}

\subsubsection{Force computations and calibration of the force coefficients}

% Once we verified the flat version of the database is able to reproduce both the surface elevation and the kinematics reasonably well, we proceed with computing the force. 
The kinematics database's purpose is to provide a reliable source of extreme wave kinematics. The choice of the mass and inertia coefficients are left to the user of the database.
The choice of $C_M$ and $C_D$ for industrial applications is usually based on best practices documents \cite{veritas2007recommended}. Another well-cited theory is the asymptotic theory from \citet{Bearman1986}, which computes the 2D theoretical values of $C_M$ and $C_D$ for low Keulegan-Carpenter numbers ($KC \rightarrow 0$).
Longoria et al. \cite{Longoria1993} used an oscillatory irregular flow on a submerged cylinder in a water tunnel to investigate the effect of irregular waves on the mass and drag coefficients of circular cylinders. For a ratio of $Re/KC \approx 2300$, they report a $C_M=1.6$ and a $C_D=1.0$ for a $KC \approx 3.0$. Some authors correct the values of the mass and drag coefficient on a per-wave basis \cite{Sumer2006}, or according to the depth below the free surface where the flow is usually more inertia-dominated.

When experimental data are available, it is possible to find the $C_M$ and $C_D$ coefficients that describe the measured forces best and facilitate the comparison of the load statistics.
\citet{Ghadirian2019c} used a least-square fit to compute the $C_M$ and $C_D$ that would result in the best fit between the measured and the simulated force time series. 
In this paper, we follow a similar approach.
As the low $KC$ number at the free surface suggests that the process is inertia-dominated for the whole cylinder, we expect the main population of the force crest distribution to depend linearly on the mass coefficient $C_M$. We also expect the drag coefficient $C_D$ to play an important role only for the large force events, in the tail of the force crest distribution. Therefore, we scale the mass coefficient $C_M$ to match the standard deviation of the experimental force signals.

To compute the optimal mass coefficient, we use the following procedure.
First, we compute the horizontal force time series using $C_M=2.0$ as first-guess value, with $C_D=1.0$. Then we derive a correction factor $\chi$ as the ratio between the standard deviation of the experimental force signal $\sigma_F^{exp}$ and the standard deviation of the experimental force signal $\sigma_F^{sim}$, by which we find a new mass coefficient $C_M^* = C_M \cdot \chi$. 
We finally rerun the force computation.

We use a constant $C_D=1.0$, following the suggestion of \citet{schloer2017experimental}. In principle, it is possible to modify the $C_D$ until a good agreement is found in the tail of the force crest exceedance probability plot. 
However, some extreme force peaks ($P<2\e{-2}$) are associated with slamming forces by breaking waves. Since the current model does not include any slamming contribution, compensating by enhancing the drag coefficient would lead to an unphysical overestimation of the drag force. 

\begin{figure}
  \centering
    \begin{subfigure}{0.9\textwidth}
      \centering
      \includegraphics[width=0.95\textwidth]{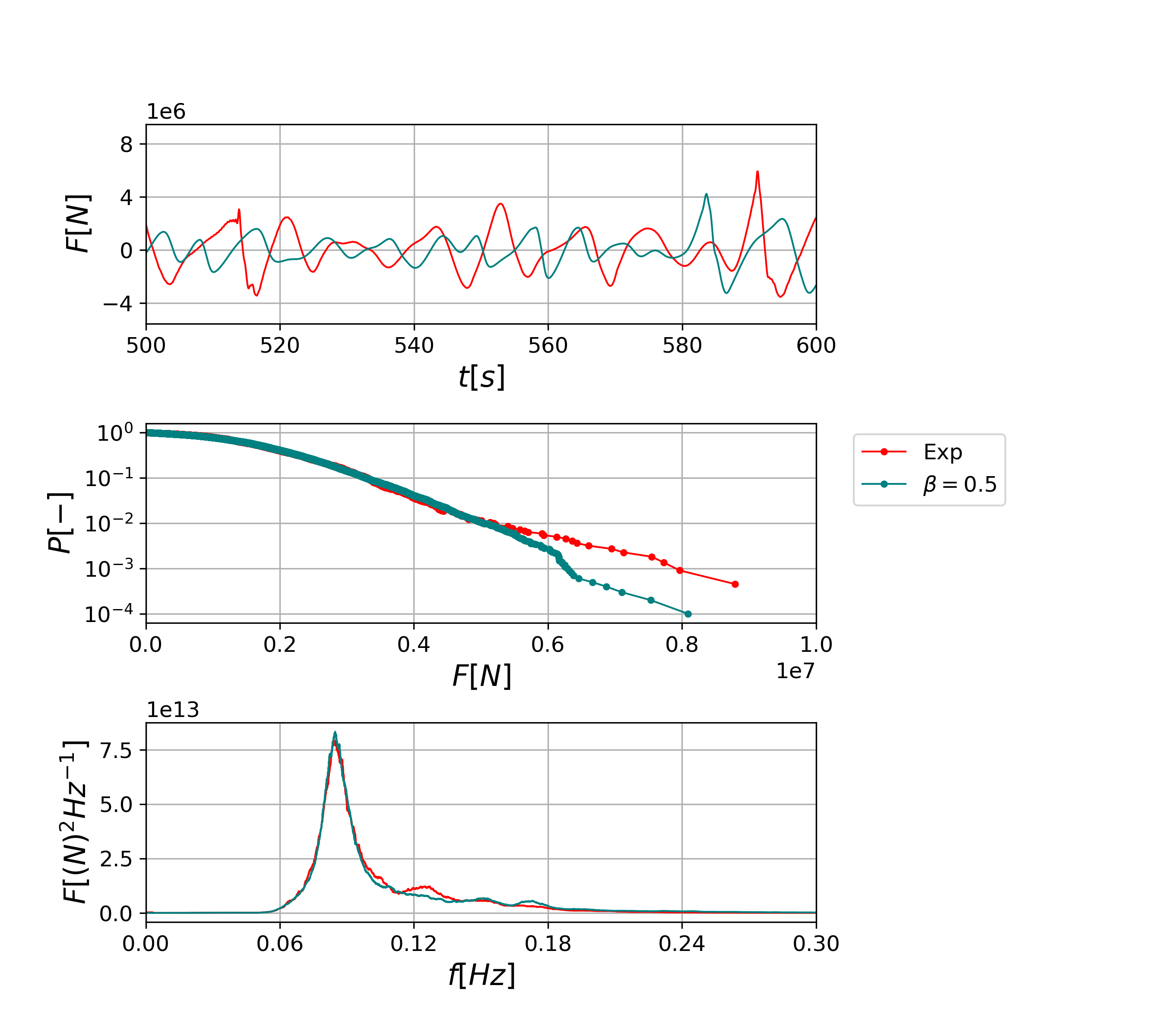}
      \caption{Case 11, 100-yr at $h=33.0[m]$. Corrected $C_M$ and $C_D$.}
      \label{fig:FFlat11_corrected}
    \end{subfigure}
  \end{figure}
  \begin{figure}
    \ContinuedFloat
    \begin{subfigure}{0.9\textwidth}
      \centering
      \includegraphics[width=0.95\textwidth]{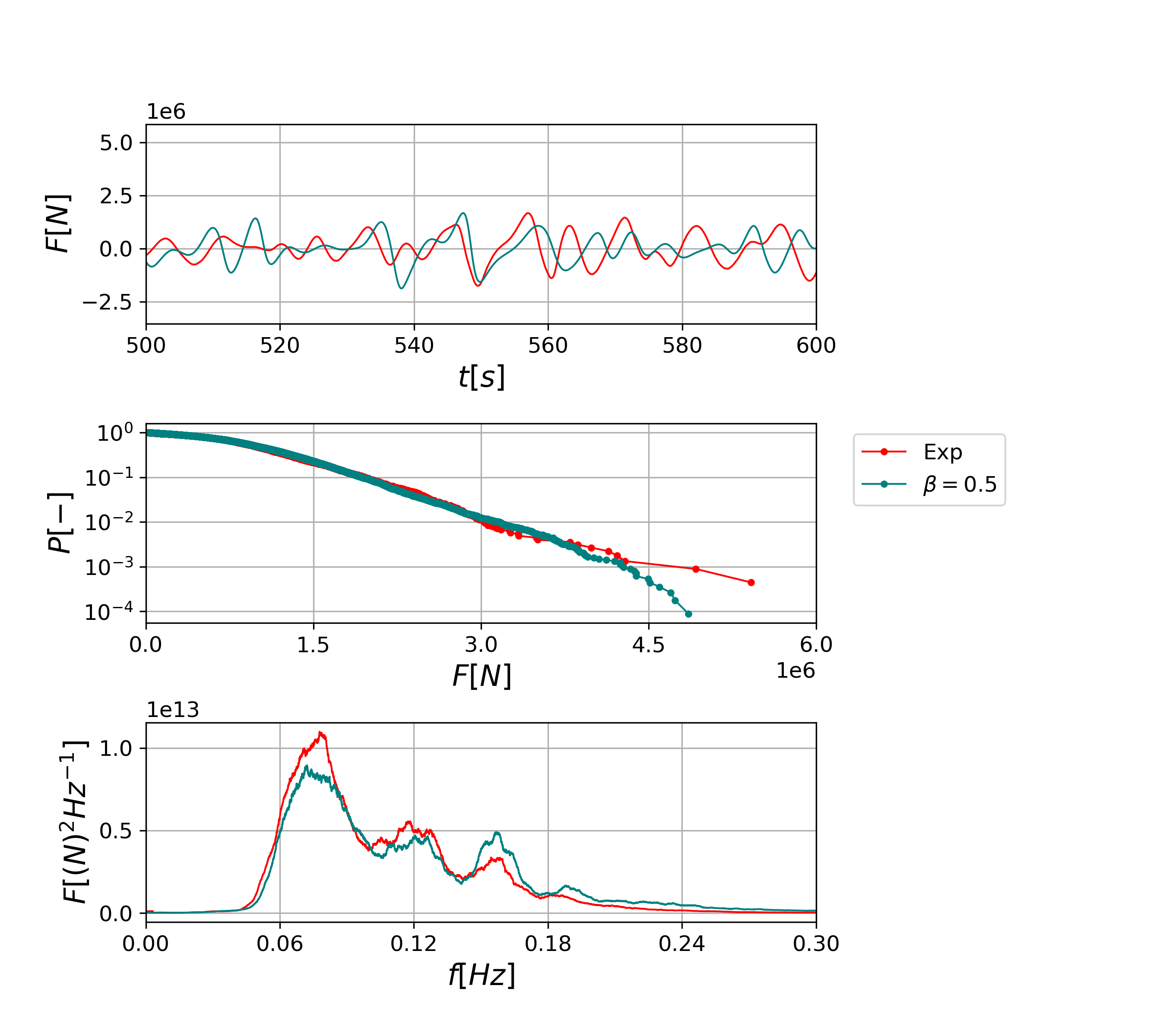}
      \caption{Case 23, 100-yr at $h=20.0[m]$. Corrected $C_M$ and $C_D$.}
      \label{fig:FFlat23_corrected}
    \end{subfigure}    
    \caption{Force computations for case 11 and 23.}
\end{figure}

We obtain a list of corrected $C_M^*$ values as summarized in \autoref{tab:gBreak}. The investigations of \citet{Longoria1993} predicted a  $C_M\in [1.6,1.8]$, comparable with the current findings. 
This approach for finding $C_M$ was rather insensitive to the choice of the drag coefficient $C_D$. When recomputing the mass coefficients with a lower drag coefficient, closer to the value suggested by Longoria et al. ($C_D=0.5$), we obtained very similar mass coefficients, with a maximum deviation of $5\%$ from the values in \autoref{tab:gBreak}.

For test 11, in \autoref{fig:FFlat11_corrected}, a very good correlation on the peak forces for up to $P\ge 1.0 \e{-2}$ is found. In the tail of the distribution, the simulations underestimate the experimental force. Part of this discrepancy is due to the absence of a slamming load model in the force computations. \citet{Pierella2019Malta} showed that coupling a slamming force model to the database allows reconstructing the extreme events with greater accuracy.
In the region between $0.1[Hz]$ and $0.2[Hz]$, we observe some secondary peaks in the experimental force spectrum. As above mentioned, this is believed to be the effect of free waves induced at the linear generation boundary.
For test 23, in \autoref{fig:FFlat23_corrected}, a similarly good agreement is achieved in the main population of the force crest peak distribution, although the extreme events are better reproduced.

\section{Ultimate Load States computation: Current approach versus industrial standard procedures}

% Once we have proven the physical accuracy of the model, we proceed with showing a typical application procedure. Moreover, we compare the model with standard industry practices for calculating extreme loads on monopile foundations.
We compare the DeRisk database methodology with the Embedded stream function (ESF) method from IEC-61400-1 \cite{iec61400-1}, with the Wi-Fi JIP methodologies \cite{DeRidder2017}, and with linear and Second-Order irregular waves from \citet{Sharma1981}.

\subsection{DeRisk database}

The procedure to extract the kinematics from the DeRisk database is outlined in section \ref{sec:loadcomputation}. To compute the horizontal force for the different experiments, we use the experiment-specific mass coefficient $C_M$ in \autoref{tab:gBreak} , a drag coefficient of $C_D=1.0$, and a fluid density of $\rho=1025.0[kg\cdot m^{-3}]$.

\subsection{Embedded stream function}
\label{sec:esf}

The methodology consists of embedding a nonlinear wave in a linear background realization. At the embedment time $t_0$, the kinematics of a linear sea state and of a stream function wave with period $T$ are blended according to a function $g(t)\in[0,1]$, where $g(t)=0$ for $t\le T/2-t_0$ and $t\ge T/2+t_0$, and $g(t)=1$ for $t=t_0$. The force is computed according to the new kinematics, and the response of the structure is simulated to extract the largest load.

This method has three main parameters: the embedment time $t_0$, the period of the stream function wave $T$ and the height of the stream function wave $H$. 
Depending on their combination, we can identify two main embedment strategies from the literature.

In the \textit{hard embedment} several stream function waves are generated by using

\begin{align*}
   H=1.86H_{S,50}   \\
   11.1\sqrt{(H_S/g)} \le T \le 14.3\sqrt{(H_S/g)}
   \label{eq:streamfunc_parameters}
\end{align*}

where $H_{S,50}$ is the significant wave height for a sea state with 50-yr return time.
They are subsequently embedded in a random time instant $t_0$ in the background regular wave realization, which is usually 3-hr or 1-hr long. When analyzing a stiff structure, a single regular stream function wave can be used.
This methodology does not retain the randomness in the height of the largest wave in a realization, which is decided a priori based on the Rayleigh distribution.
The constrained wave methodology has further limitations when a flexible structures is analyzed. In this case, the stochastic nature of the load history is important as the static and dynamic loads can freely combine, and the maximum load on the structure does not necessarily happen when the largest wave passes by.
This shortcoming is partially addressed by the \textit{soft embedment} methodology, where the largest wave of the realization is suitably   replaced by a nonlinear stream function wave \cite{rainey2007weak, Shaofeng10MW2020}.
% and give different results than when only the static loads are included. If one were to include the stochasticity of the loads, we would observe that
In this work, we consider a stiff cylinder and hard-embed one single cycle of a regular stream function wave with three different periods $T$ in the aforementioned interval and $H=1.86 H_S$.

\subsection{The WiFi industry project}

The WiFi Joint Industry Project (JIP) \cite{DeRidder2017} provides a method to compute the extreme force with a probability of occurrence of 1/1000 in a 1-hr sea state. 
First, an average steepness parameter is calculated as $s_P=H_S/L_P$, where $L_P$ is the length of the linear wave associated with the spectral peak period. If the steepness $s_P\ge 0.04$, the largest force is assumed to come from a large breaking wave, and the slamming load $F_B$ can be computed as

\begin{align}
  F_B=\frac{1}{2}\rho C_S A u^2 \\
  C_S = 2\pi \\
  A = \frac{1}{32}H_B D \pi \\
  u = 1.1 \frac{L_B}{T_B}
  \label{eq:wifi}
\end{align}

where

\begin{align}
  T_B=0.9T_P \\
  H_B = \min\left[1.4H_S, 0.78h,L_B\cdot\tanh\left(\frac{2\pi h}{L_B}\right)\right]
\end{align}

In the above equations, $\rho$ is the density of the fluid, $C_S$ is the slamming coefficient by \citet{goda1966study}, while $D$ is the cylinder diameter. If $s_P < 0.04$, the load is computed by using the standard Morison equation.
The impulsive slamming load is to be applied on top of the quasi-static background load when the wave crest first touches the structure.

\subsection{Linear and Second-order irregular waves}

Another widely used methodology to simulate nonlinearity in extreme sea states is the second-order irregular wave theory by \citet{Sharma1981}, which accounts for a mild nonlinearity in the wave kinematics together with the stochastic nature of the process.
Here, 12 different 1-hr realizations of the same sea state are performed and stacked together to obtain a 12-hr total time series. 
To generate the spectrum, a return period of $T_{\text{ret}}=3600.0[s]$ was used. The high cut-off of the spectrum was $f_{\text{hc}}=1/3 [Hz]$, consistently with the OceanWave3D simulations, while the time domain signal resolution was $\Delta t=0.5 [s]$.

A total of $n_z=20$ points was used to compute the wave kinematics, evenly distributed from the sea bottom to the still water line (SWL). We use both the first-order and the first- plus second-order kinematics to compute the horizontal force on the stiff cylinder via the Rainey force model. The kinematics are extrapolated above SWL consistently up to second-order \cite{veritas2007recommended}. The first-order kinematics is linearly extrapolated up to the first-order wave elevation, while the second-order kinematics is integrated only up to the SWL.

\vspace{1em}

\subsection{Comparison}

To compare the results, we have chosen the two significant wave heights associated with tests 11 and 23 from \autoref{tab:derisk_exp}, corresponding to 100-yr return sea states.
% The following results are computed using the sloped version of the database, with the methodology described \autoref{sec:loadcomputation} and experiment specific $C_D$ and $C_M$ as from \autoref{tab:gBreak}.

\begin{figure}[htbp]
  \centering
  \includegraphics[width=0.9\columnwidth, clip, trim={0 7.5cm 0 0}]{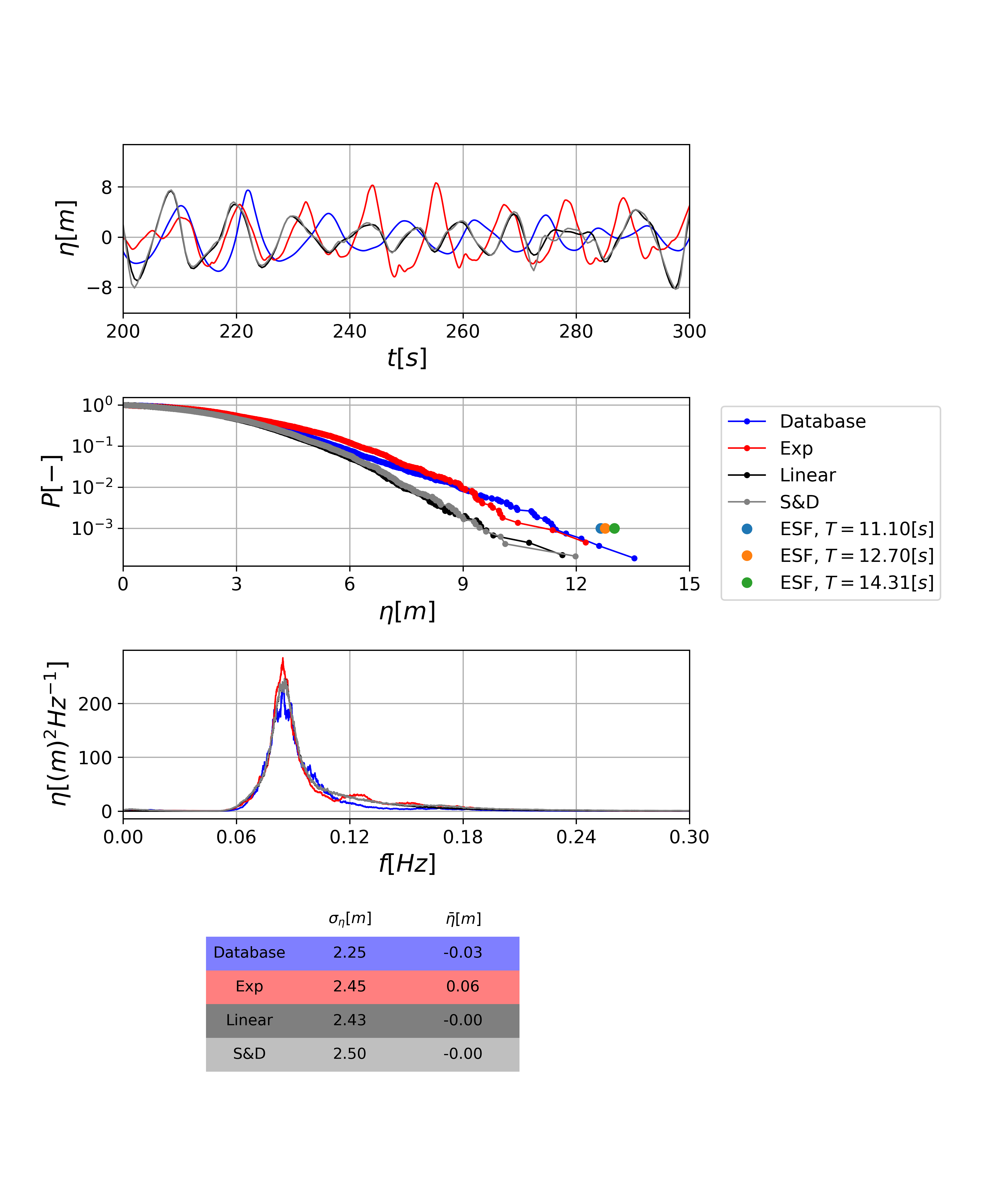}
  \caption{Experiment 11. Comparison of the experimental free surface elevation with the design methodologies.}
  \label{fig:exp11_comparison_eta}
\end{figure}

\begin{figure}[htbp]
  \centering
  \includegraphics[width=0.9\columnwidth, clip, trim={0 7.5cm 0 0}]{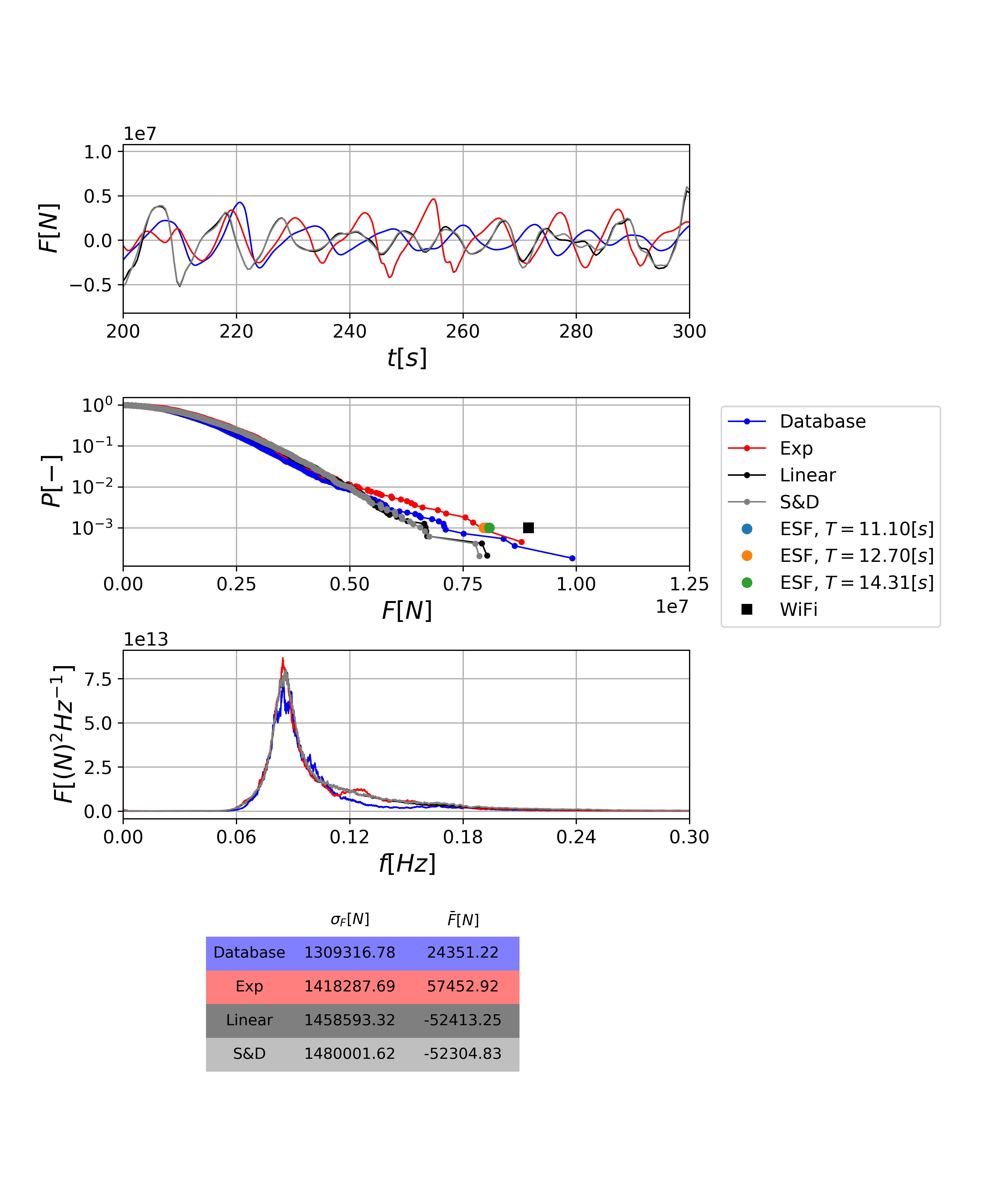}
  \caption{Experiment 11. Comparison of the experimental inline force with the design methodologies.}
  \label{fig:exp11_comparison_F}
\end{figure}

In \autoref{fig:exp11_comparison_eta}, the modelled free surface elevation signals for experiment 11 are compared.
In black, twelve 1-hr long linear realizations were combined to obtain the linear computation.
In gray, the Sharma and Dean second-order contribution was added to the linear one.
Five different computations were extracted from the database, which, after disregarding the transients and applying the scaling procedure, yields a time series of roughly 16 hours, in blue. 

In \autoref{fig:exp11_comparison_eta}(b), the main population of the samples in the database crest distribution is in line with the experimental results, down to the extreme crest heights for $P=1.0\e{-3}$. On the other hand, the largest ESF wave crest is conservative. 
As for the power spectra, we notice a satisfactory agreement across the frequency span for all of the methods. The Sharma and Dean solution matches the experimental superharmonic energy distribution best, but is the least accurate in predicting the free surface elevation crest distribution.

As for the force crest distribution, in \autoref{fig:exp11_comparison_F}, all of the methods perform fairly well down to a probability of $10^{-2}$.
The ESF method is able to catch the experimental force very well, while the WiFi prediction is slightly conservative. The other methods underestimate the force peak by $20\%$. This is expected, as by analysis of the force time series it was seen that the experimental event with $P=1\e{-3}$ was associated with a breaking event inducing a slamming load.

\begin{figure}[htbp]
  \centering
  \includegraphics[width=0.9\columnwidth, clip, trim={0 7.5cm 0 0}]{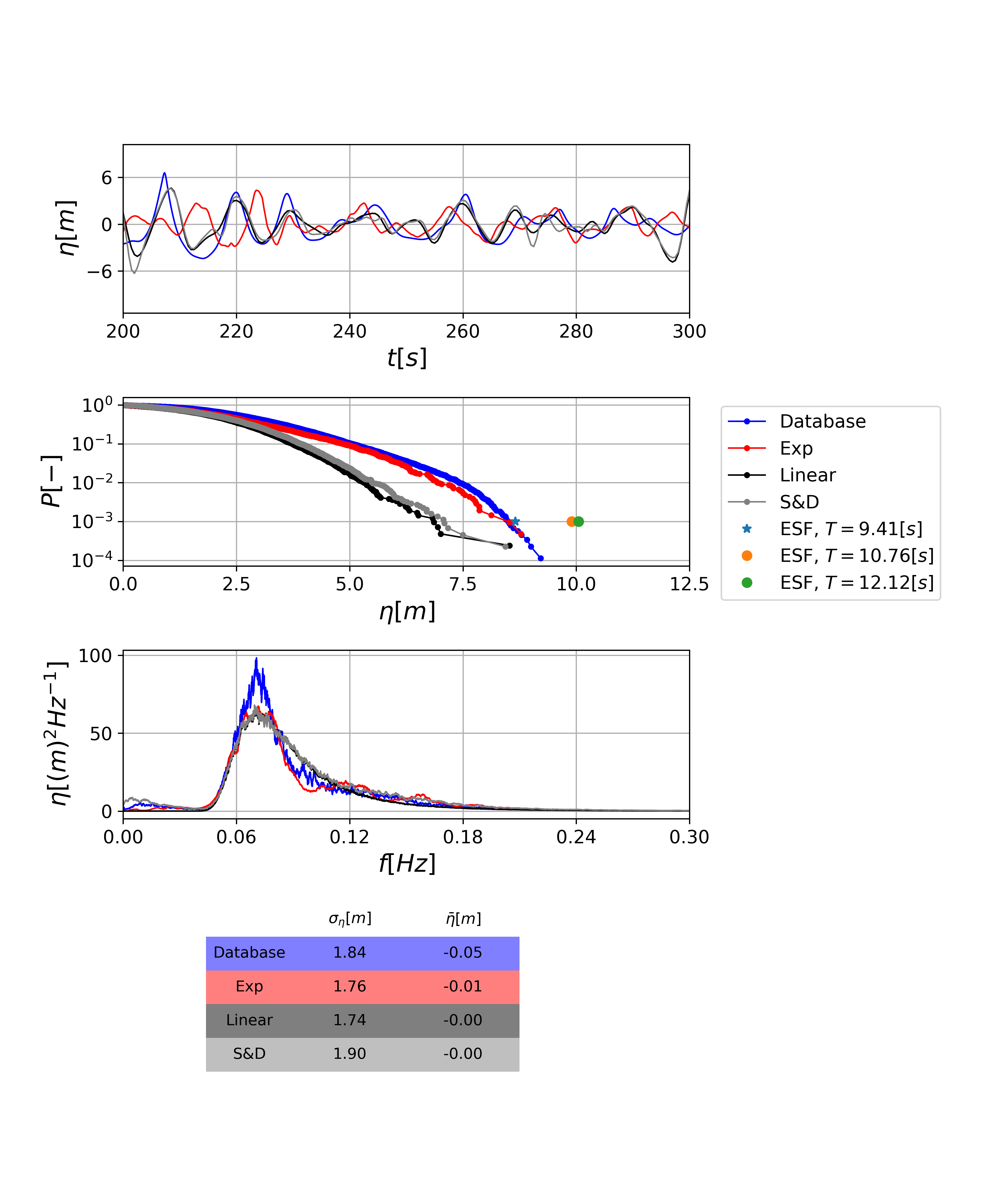}
  \caption{Experiment 23. Comparison of the experimental free surface elevation with the design methodologies.}
  \label{fig:exp23_comparison_eta}
\end{figure}

\begin{figure}[htbp]
  \centering
  \includegraphics[width=0.9\columnwidth, clip, trim={0 7.5cm 0 0}]{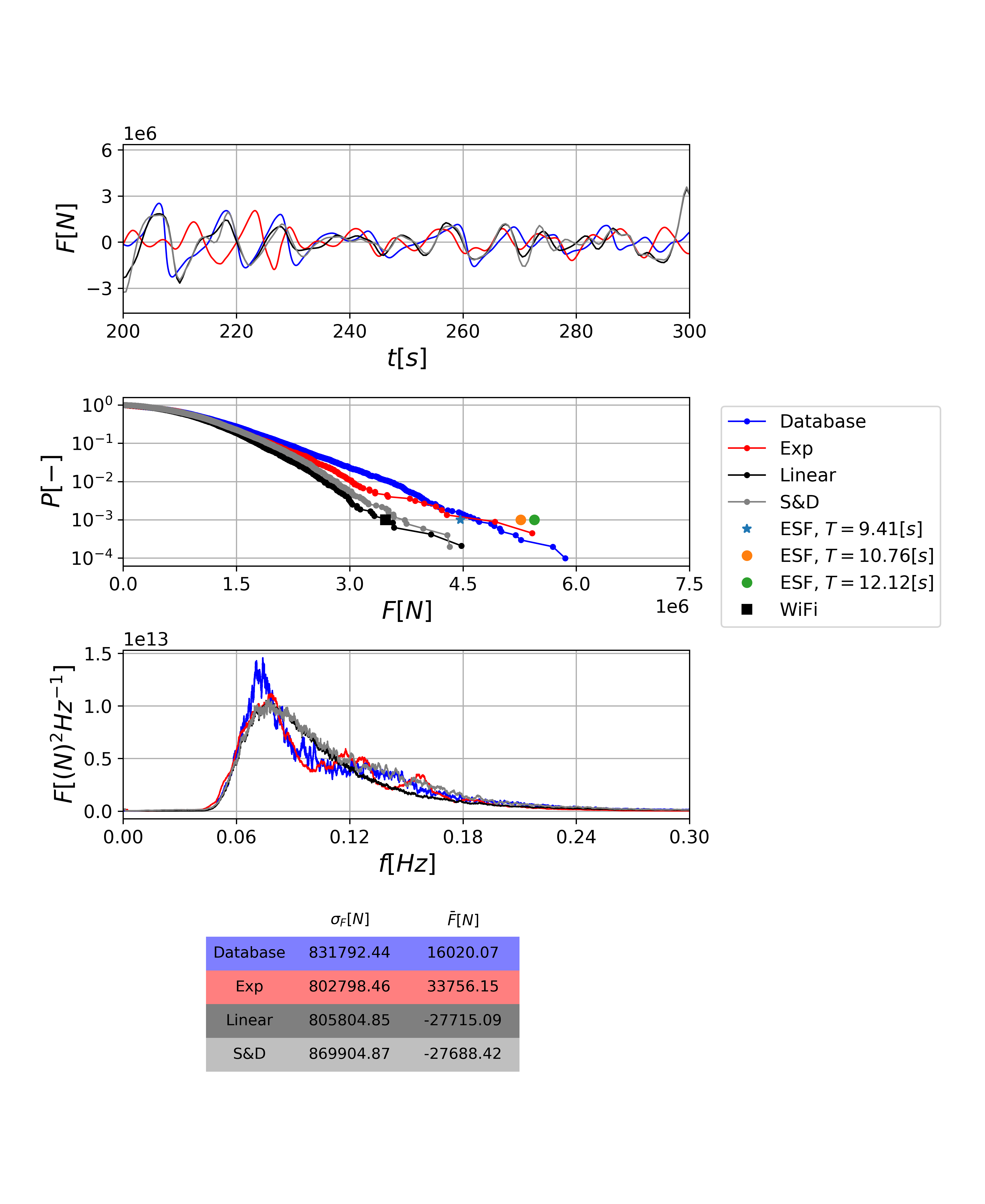}
  \caption{Experiment 23. Comparison of the experimental inline force with the design methodologies.}
  \label{fig:exp23_comparison_F}
\end{figure}

In \autoref{fig:exp23_comparison_eta}, we observe the results for a 100-yr storm in a shallower depth $h=20.0[m]$.
For a $P=1\e{-3}$, the predictions from fully-nonlinear methodologies agree much better with the experiments than the linear and the Sharma and Dean solutions.
The ESF method predicts a largest crest of about $10.0[m]$ for $P=1\e{-3}$, which is rather conservative.
As for the power spectra, the superharmonic energy distribution is well predicted by the database results, although the energy at the peak frequency is overestimated. The contrary is true for the linear and the Sharma and Dean solution, which capture the peak energy but not the decay rate of energy at the superharmonic frequency.
In \autoref{fig:exp23_comparison_F}, the peak force distribution computed from the database kinematics agrees well with the experiments around the $P=1\e{-3}$ level, while the linear and the Sharma and Dean results are nonconservative.
Since the average wave steepness was $H_S/L_P=0.038$, the WiFi model did not predict that the largest load would come from a breaking event.

\begin{sidewaystable}
  \small
  \input{tables/latexetamax}
  \caption{Maximum free surface elevation $\eta_{max}[m]$, corresponding to an exceedance probability level of $P=1\e{-3}$.}
  \label{tab:maxeta}
\end{sidewaystable}

The crests with a $P=1\e{-3}$ are summarized in \autoref{tab:maxeta} for all of the cases. The picture is consistent with our previous analysis: the results from the DeRisk database are the closest to the experimental results, while the embedded stream function predictions are almost consistently conservative. The linear and the Sharma and Dean wave theories always provide too low crest estimates. 

\begin{sidewaystable}
  \small
  \input{tables/latexFmax}
  \caption{Maximum inline force, measured in $F_{max}[MN]$, corresponding to an exceedance probability level of $P=1\e{-3}$. When the maximum load for the WiFi model is associated with a breaking wave, the figure is underlined.}  
  \label{tab:maxf}
\end{sidewaystable}

In \autoref{tab:maxf} we summarize the force peaks with a $P=1\e{-3}$. For the milder storms with a return time of 10 years, both at the shallow and the deep location, all methods are quite comparable. An exception is made for the WiFi method, which is sensitive to the average sea steepness. For example, in tests 10 and 20 the breaking wave load is not triggered, contrarily to tests 1 and 9, where it leads to an overestimation of the experimental load.
For these small storms, the linear and the Sharma and Dean force predictions are acceptable, and not far from the fully nonlinear kinematics ones.

For the 100 years storms, the linear and the Sharma and Dean predictions are too nonconservative. The DeRisk database results agree better for the shallow than for the deep location. This is because in tests 11 and 12, where waves are on average steeper than in tests 22 and 23, slamming loads from breaking waves are more frequent. The ESF method performs better than the WiFi model, which seems to provide conservative estimates, as also observed by \citet{DeRidder2017}.  
As for the 1000 year return storms, none of the models is clearly superior to the others. At $h=20.0[m]$, all models output similar predictions for case 25, while for case 24 all the methods underpredict the measured forces.

The application of a slamming model in combination with the database is part of our current work. Initial results are published in \citet{Pierella2019Malta}, and
these show that the force predictions from the database can be improved if a breaking wave load is added to the quasi-static load.

% In this plot, we could not compute any force associated with the stream function wave, which breaks for $H=1.86 H_S$. This is another situation where the state-of-the art methodologies show limitations, that do not apply to the presented methodology.

%%% Local Variables:
%%% mode: latex
%%% TeX-master: "../DeRiskDB"
%%% End:

%% file: tables/derisk_exp.tex
1 &8.50 & 13.50 & 1.53 & 9.12 & 12.28 & 33.00 &  10 \\
9 &7.50 & 12.00 & 2.04 & 7.78 & 11.16 & 33.00 &  10 \\
10 &7.50 & 15.00 & 1.00 & 7.36 & 13.61 & 33.00 &  10 \\
11 &9.50 & 12.00 & 3.57 & 9.81 & 11.62 & 33.00 &  100 \\
12 &9.50 & 15.00 & 1.17 & 9.18 & 13.61 & 33.00 &  100 \\
13 &11.00 & 15.00 & 1.73 & 10.26 & 13.68 & 33.00 &  1000 \\
20 &5.80 & 12.00 & 1.02 & 6.05 & 12.10 & 20.00 &  10 \\
21 &5.80 & 15.00 & 1.00 & 6.16 & 12.41 & 20.00 &  10 \\
22 &6.80 & 12.00 & 1.58 & 7.09 & 12.25 & 20.00 &  100 \\
23 &6.80 & 15.00 & 1.00 & 7.04 & 14.06 & 20.00 &  100 \\
24 &7.50 & 15.00 & 1.00 & 7.65 & 14.06 & 20.00 &  1000 \\
25 &5.80 & 9.00 & 4.27 & 6.09 & 8.82 & 20.00 &  1000 \\

%% file: tables/latexetamax.tex
\begin{tabular}{lrrrrrrrrrrrr}
\toprule
                  & \multicolumn{12}{c}{Experiments} \\ 
\cmidrule(lr){2-13}                  
    No.         &    1&   9 &      10 &       11 &       12 &       13 &      20 &      21 &       22 &       23 &      24 & 25\\
     Ret [yr]    &    10&   10 &      10 &       100 &       100 &       1000 &      10 &      10 &       100 &       100 &      1000 & 1000\\             
     Depth [m]  & 33.0 & 33.0& 33.0& 33.0& 33.0& 33.0 & 20.0 & 20.0 & 20.0 & 20.0 & 20.0 & 20.0 \\
    $h/gT_P^2 \e{2}$ & 2.23 & 2.70 & 1.82 & 2.49 & 1.82 & 1.80 & 1.39 & 1.32 & 1.36 & 1.03 & 1.03 & 2.62 \\
    $H_S/gT_P^2\e{3}$ & 6.17 & 6.37 & 4.05 &7.41 &5.05 &5.59 &4.21 &4.08 &4.82 &3.63 &3.95 &7.98 \\
\midrule
Exp &   10.54  &   9.70  &   8.71  &   11.37  &   10.73  &   11.59  &   6.76  &   7.33  &    7.64  &    8.53  &   9.47  &   6.73  \\
\midrule
Database  &   10.86  &   9.02  &   9.37  &   11.46  &   10.76  &   11.54  &   6.85  &   7.51  &    8.38  &    8.50  &   8.69  &   6.60  \\
Linear    &   8.92  &   7.28  &   6.77  &    9.61  &    8.94  &   10.24  &   5.62  &   5.73  &    6.53  &    6.86  &   7.45  &   5.49  \\
S\&D      &    8.68  &   7.37  &   7.03  &    9.44  &    8.75  &    9.95  &   5.97  &   6.15  &    7.05  &    7.09  &   7.69  &   5.50  \\
ESF       &   11.73  &   9.26  &   8.65  &   13.01  &   11.85  &   13.86  &   7.85  &   8.29  &   10.16  &   10.05  &   9.25  &   8.15  \\
\bottomrule
\end{tabular}

%% file: tables/latexFmax.tex
\begin{tabular}{lrrrrrrrrrrrr}
\toprule 
            & \multicolumn{12}{c}{Experiments} \\ 
\cmidrule(lr){2-13}                
No.         &    1&   9 &      10 &       11 &       12 &       13 &      20 &      21 &       22 &       23 &      24 & 25\\
Ret [yr]    &    10&   10 &      10 &       100 &       100 &       1000 &      10 &      10 &       100 &       100 &      1000 & 1000\\             
Depth [m]  & 33.0 & 33.0& 33.0& 33.0& 33.0& 33.0 & 20.0 & 20.0 & 20.0 & 20.0 & 20.0 & 20.0 \\
$h/gT_P^2 \e{2}$ & 2.23 & 2.70 & 1.82 & 2.49 & 1.82 & 1.80 & 1.39 & 1.32 & 1.36 & 1.03 & 1.03 & 2.62 \\
$H_S/gT_P^2\e{3}$ & 6.17 & 6.37 & 4.05 &7.41 &5.05 &5.59 &4.21 &4.08 &4.82 &3.63 &3.95 &7.98 \\
\midrule
Exp & 5.88 & 5.66 & 4.45 & 7.96 & 7.52 & 9.21 & 3.41 & 3.56 & 4.29 & 4.92 & 6.72 & 3.32 \\
\midrule
Database    & 6.06 & 5.03 & 4.60 & 7.12 & 6.08 &  7.17 & 3.39 & 3.96 & 4.95 & 4.70 & 4.94 & 3.35 \\
Linear      & 5.37 & 4.74 & 4.23 & 6.67 & 5.41 &  6.90 & 3.26 & 3.19 & 3.84 & 3.47 & 3.97 & 3.62 \\
S\&D         & 5.27 & 4.69 & 4.23 & 6.56 & 5.35 &  6.72 & 3.42 & 3.32 & 4.18 & 3.73 & 4.28 & 3.67 \\
WiFi        & \underline{9.55} & \underline{8.06} & 4.23 & \underline{8.95} & \underline{9.57} & \underline{10.49} & 3.26 & 3.19 & \underline{5.37} & 3.47 & \underline{5.89} & \underline{3.71} \\
ESF         & 6.78 & 5.23 & 4.81 & 8.08 & 6.92 &  9.30 & 4.22 & 4.31 & 5.82 & 5.45 & 4.94 & 4.09 \\
\bottomrule
\end{tabular}

%% file: texFiles/conc.tex
\section{Conclusion}

In this work, we established a database of nonlinear wave kinematics for ultimate load state computations. Different realizations of nonlinear waves are easily extracted from the database, and coupled with a force model to compute loads on slender offshore structures.
We demonstrated that the validity of the database is extended by applying the Froude hypothesis to scale the database kinematics to match the required application scale.

Through computations on a flat bed domain, we investigated the effect of the breaking filter on the crest statistics. By comparison with experimental data, we found that a value of $\beta=0.5$ gave the best agreement.
When using the nonlinear kinematics to reproduce the horizontal force from experiments on a stiff monopile, a good agreement was observed. Some discrepancies in the super-harmonic part of the horizontal force spectrum were present. These may result from using linear wave generator theory in the experiments.

The static loads from state-of-the-art design methods were compared to the present method.
For milder storms, all the analyzed methods output acceptable results and predicted the experiments well. For larger storms, the database results were consistently better than the linear and the Sharma and Dean predictions, while they generally agreed with the embedded stream function predictions. 
The WiFi JIP force model output accurate but slightly conservative predictions at $33.0[m]$ depth, while it was generally nonconservative at $20.0[m]$ depth.

In the next phase of the DeRisk project, the database will be extended to directional seas. Further work will also be done to optimize the procedure outlined in \citet{Pierella2019Malta} to compute slamming loads directly from the database results by using the pressure impulse model by \citet{ghadirian_bredmose_2019}.